\documentclass[10pt]{article}
\usepackage{amsmath}
\usepackage{graphicx}
\usepackage{microtype}
\usepackage{textcomp}
\usepackage{algorithmic}
\usepackage{tikz}
\usepackage{hyphenat}
\usepackage{epsfig,psfrag,amstext,amsfonts,amssymb,subeqnarray,color}
\usepackage{amsthm,algorithmic}
\usepackage[linesnumbered,ruled]{algorithm2e}
\usepackage[caption=false,font=footnotesize,subrefformat=parens,labelformat=parens]{subfig}
\usepackage[framemethod=tikz]{mdframed}
\usepackage{url}
\usepackage[hyperindex,pdftex]{hyperref}
\usepackage{epstopdf,bm,multirow}
\usepackage{spconf}
\newmdenv{eqnbox}

\DeclareMathOperator*{\argmin}{arg\,min}

\title{FedLoc: Federated Learning Framework for Data-Driven Cooperative Localization and Location Data Processing}

\name{Feng Yin$^{*1}$, Zhidi Lin$^{1}$, Yue Xu$^{2}$, Qinglei Kong$^{1}$, Deshi Li$^{3}$, Sergios Theodoridis$^{1}$, Shuguang (Robert) Cui$^{1}$}
\address{1. The Chinese University of Hong Kong (Shenzhen) and SRIBD, 518172, China \\ 2. BeiJing University of Post and Telecommunications \\ 3. Wuhan University}

\begin{document}
\maketitle
\begin{abstract}
In this overview paper, data-driven learning model-based cooperative localization and location data processing are considered, in line with the emerging machine learning and big data methods. We first review (1) state-of-the-art algorithms in the context of federated learning, (2) two widely used learning models, namely the deep neural network model and the Gaussian process model, and (3) various distributed model hyper-parameter optimization schemes. Then, we demonstrate various practical use cases that are summarized from a mixture of standard, newly published, and unpublished works, which cover a broad range of location services, including collaborative static localization/fingerprinting, indoor target tracking, outdoor navigation using low-sampling GPS, and spatio-temporal wireless traffic data modeling and prediction. Experimental results show that near centralized data fitting- and prediction performance can be achieved by a set of collaborative mobile users running distributed algorithms. All the surveyed use cases fall under our newly proposed \textit{Federated Localization (FedLoc)} framework, which targets on collaboratively building accurate location services without sacrificing user privacy, in particular, sensitive information related to their geographical trajectories. Future research directions are also discussed at the end of this paper.
\end{abstract}

\begin{keywords}
Cooperation, Data-driven models, Distributed processing, Federated learning, Gaussian processes, Location services, User privacy. 
\end{keywords}

\section{Introduction}
\label{sec:introduction}
With the explosion of data and the ever-increasing computing power, we have witnessed nowadays the popularity of machine learning models and algorithms which are data-driven. In principal, with more data, an underlying complex system/dynamic/regression function can be closely approximated. However, when the data size increases beyond a limit, both the scale of the model and the computational complexity of an associated learning algorithm can become computationally tough. For instance, the computational complexity for training a Gaussian process model scales cubically with the data size \cite{RW06}. This renders the required computational load for sophisticated data-driven learning models prohibited for practical cases. 

The recently proposed federated learning framework \cite{mcmahan17a} has received a lot of attention, as it enables a large-scale machine learning models to be trained jointly by a large number of mobile users through cooperation. Actually, there exist various similar works before the federated learning, for instance \cite{Povey15, Neverova16}, but federated learning emphasizes more on the following aspects: (1) non-i.i.d. data; (2) unbalanced local data size; (3) large number of local users; (4) limited communication; and (5) data privacy \cite{mcmahan17a}. It deserves to highlight that federated learning is a promising technical solution to solve the ever-increasing concerns about the loss of user privacy and to meet the ever-stringent data protection regulations world-wide, for instance, the General Data Protection Regulation (GDPR) implemented by the European Union in 2018. Federated learning has triggered various potential applications in the sectors of smart medicine, finance, and next-generation wireless communications {\cite{tran2019federated, samarakoon2018federated,lee2018privacy}}. In this paper, we extend federated learning to a new application sector, namely target localization and location-related services. 

Target localization is meant to provide an estimate of the desired position as accurate as possible. There exist a plethora of state-of-the-art techniques for static target localization, target tracking, navigation, and interested readers can refer to \cite{GG05, Sayed05, Bar-Shalom2001} and the references therein for more information. Most of these techniques rely on empirical, parametric transition and measurement models, which can be regarded as an individual abstract of human experience, thus they may severely mismatch the underlying mechanism in complicated environments such as office, shopping mall, museum, etc. However, directly learning from a huge volume of historical data may help alleviating such a model mismatch and improve the positioning accuracy even further. 
Apart from the traditional localization service, a new type of location related services have emerged in the recent years under the umbrella of smart cities, namely the spatio-temporal location data prediction. This type of services include, but not limited to, wireless traffic prediction, taxi supply and demand prediction, energy consumption prediction, air pollution prediction at specific locations. Data-driven, learning model-based solutions have demonstrated great data representation and generalization capability {\cite{Xu19,liu2019co,kim2019pre,qi2018deep}}. 

However, the greatest difficulty that we confronted when applying machine learning models to localization and location data modeling lies in the big amount of labeled training data, which can be solved by aggregating small data collected from a large number of mobile users. Yet, such data gathering processes may cause severe data privacy issues, particularly when location is involved. As a special example, during the COVID-19 pandemic we have seen the value of sharing trajectories to track the spread of infections and predicting high-risk regions, meanwhile, there is an urgent need for location privacy preservation of the mobile users \cite{COVID-19}. The federated learning framework is an outstanding solution for enhancing wireless localization accuracy and maintaining safe cooperation among users at the same time. 


The gist of the proposed Federated Localization (FedLoc) framework is to let each mobile user/smart agent collect a smaller scale, local data set and approximate the global machine learning model in a cooperative manner. Some concrete examples are as follows: (1) For static localization, a number of mobile users collect radio features at specific positions obtained either from the global positioning system (GPS) (for outdoor scenarios) or from the proximity to indoor reference points/landmarks (for indoor scenarios); (2) For target tracking and navigation, the mobile users collect diverse trajectories of inertial sensor- and wireless observations; (3) For wireless traffic prediction, base stations work as smart agents to collect local wireless data usage generated by their serving mobile users. We believe that the FedLoc framework is an up-and-coming solution for futuristic data-driven cooperative localization, not only because of the rapid development of distributed optimization techniques that serve as the algorithmic core, but also largely owing to the rapid development of smart phones with ever-increasing computation power and network throughput, the widespread use of quick-response (QR) codes, and the high-precision indoor/outdoor maps, altogether. Therefore, we believe it is timely to exploit all relevant federated learning techniques for localization and location data processing.    

This overview paper is a four-mode mixture of review, new proposals, real evaluations, and outlook, being different from the majority that solely review the existing works. We focus on a specific application sector of federated learning, namely the data-driven cooperative localization and location data processing. The models and algorithms to be reviewed are carefully tailored for our desired applications. Besides, we focus on real use cases and their practical implementations from our own works as well as some other related works that all fall under this new cooperative paradigm. Detailed contributions of this overview paper are as follows. 
\begin{itemize}
\item First, we propose a federated localization framework, {called} FedLoc, which elegantly addresses the privacy issue in cooperation among a massive number of mobile users for target localization and location data processing. We also proposed two potential wireless network infrastructures, namely a cloud-based one and an edge-based one, that can potentially help meet the communication requirements of the FedLoc framework.
\item Second, we clarify the differences between the proposed FedLoc framework and the existing cooperative localization framework for sensor networks as well as the classic crowd-sourcing framework. 
\item Third, we review some state-of-the-art federated learning procedures, two widely used learning models, namely the deep neural network (DNN) and Gaussian process (GP), and a few distributed model hyper-parameter optimization schemes that work reasonably well for the two learning models. We put more emphasis on the Bayesian GP models than then deterministic DNN models due to their unique welcome features for modeling location data.
\item Fourth, we discuss four concrete use cases, namely (1) static target localization/fingerprinting; (2) outdoor vehicle navigation; (3) indoor pedestrian tracking; and (4) spatio-temporal wireless traffic prediction, to explain the use of the FedLoc framework. In the first use case, a static target localization system is built based on a DNN that maps a vector of radio features to a desired position. In the second use case, we propose a DNN-based accurate vehicle navigation with low-sampling-rate GPS. In the third case, the state transition function, as represented by the GP model, maps the current state to the next state in a non-parametric way for indoor pedestrian motion modeling. In the fourth use case, wireless traffic is modeled by a scalable GP under 5G Cloud-Radio Access Network (C-RAN) infrastructure. Various other related applications are also mentioned in this paper. 
\item Lastly, we evaluate the proposed FedLoc framework with real data sets for two aforementioned use cases to demonstrate their practical implementations and effectiveness in reality.
\end{itemize}

In this overview paper, we concentrate on federated learning tailored to target localization and location data processing. Due to the space limitation as well as the expertise of the authors, the following aspects are only briefly touched upon. 
\begin{itemize}
\item Distributed optimization methods in the contexts of robustness, communication efficiency, and low-complexity. Some recent works include \cite{guha2019oneshot, Sattler19}.
\item Adversarial attacks and advanced privacy-preserving schemes such as the block-chain based ones for federated learning. Some recent works include \cite{bhagoji2018analyzing, bagdasaryan2018backdoor, kim2018blockchained}.
\item General techniques and challenges of federated learning as well as its applications in other industry sectors, as surveyed by \cite{Yang19, li2019federated}. 
\end{itemize}

\begin{figure}[t]
	\includegraphics[width=0.48\textwidth]{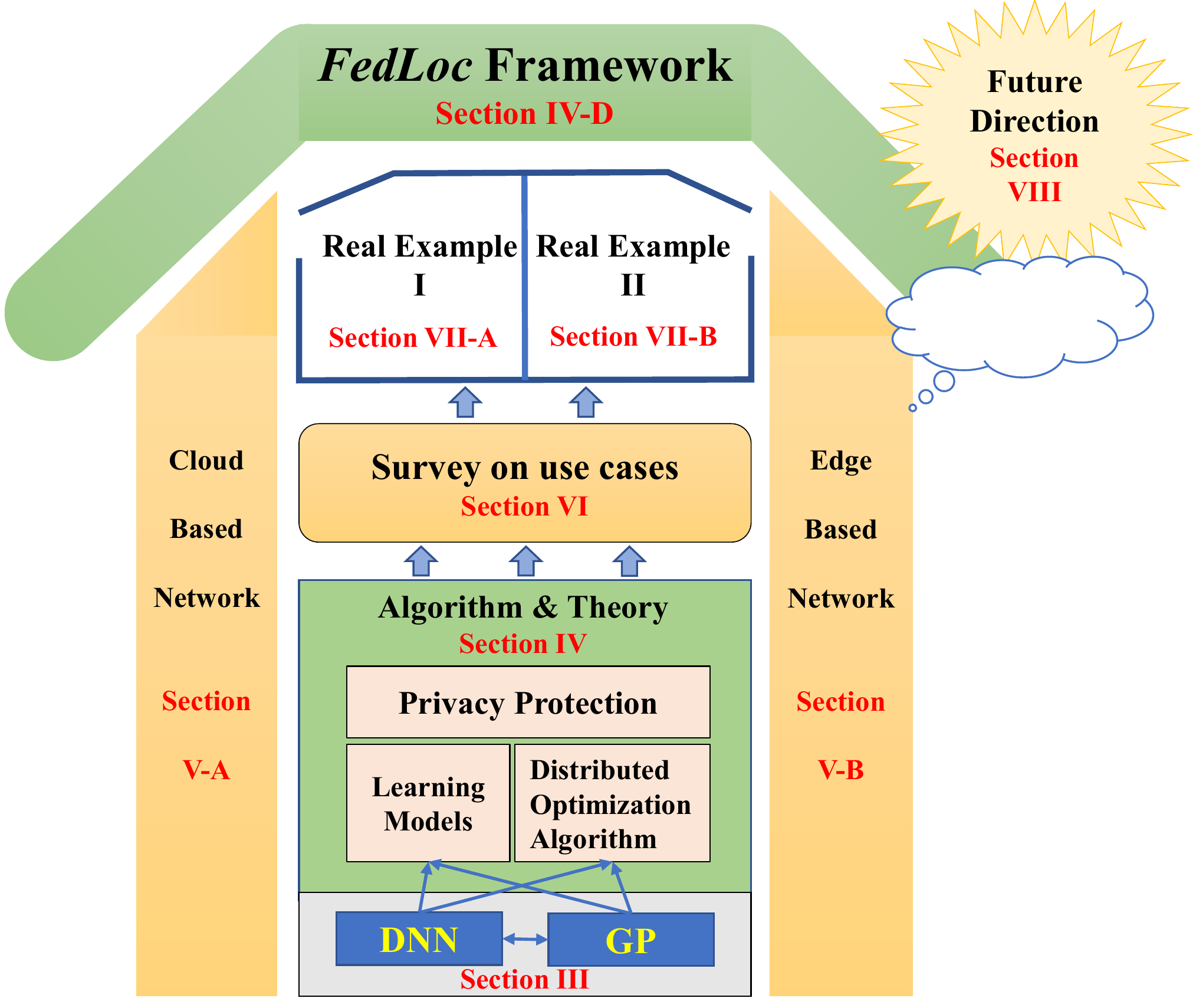}
	\caption{Overall organization of this paper and links between different sections.}
	 \label{fig:frametwork_all}
\end{figure}

The rest of this paper is organized as follows. In Section~\ref{sec:relatedwork}, we briefly review the existing ``cooperation" frameworks proposed primarily for wireless sensor networks. In Section~\ref{sec:lmodels}, we introduce two important learning models, namely the deep neural network and Gaussian process, for learning from data. In Section~\ref{sec:fwp}, we introduce the proposed FedLoc framework in detail, followed by two different wireless network infrastructures given in Section~\ref{sec:NetInf} to support the real deployment of the FedLoc framework. Various use cases of the proposed FedLoc framework are showcased in Section~\ref{sec:use-cases}. Simulation results are given in Section~\ref{sec:results} to demonstrate the effectiveness of the FedLoc framework. In Section~\ref{sec:challenges}, we discuss the major challenges of the FedLoc framework and give a few future research directions. Lastly, Section~\ref{sec:conclusion} concludes this paper. Figure~\ref{fig:frametwork_all} gives a clear global picture of our work.

\section{Related Work}
\label{sec:relatedwork}
In this section, we survey all related works and clarify their differences from our FedLoc framework to be introduced in Section~\ref{sec:fwp}. 
\subsection{Sensor Network Localization}
When speaking of ``cooperation'' in the context of wireless localization, it will certainly remind us the class of algorithms for determining a number of agents (nodes with unknown positions) with the aid of a few anchors (nodes with known positions) and a bunch of wireless measurements made between these nodes. 

Cooperative localization has gained much attention since 2005 owing to the seminal work by Patwari and Hero \cite{PAK05}, where they proposed to use the simple least-squares estimation criterion with time-of-arrival (ToA) or received-signal-strength (RSS) measurements to localize dozens of agents. The proposed method was evaluated with two sets of real measurements collected in an indoor environment. This seminal work has triggered a plethora of methods in the following years. Representative works include \cite{BLWY06, WLW09, Shen10, Win11, Yin15b, Jin20}, to mention a few.


The fundamental differences between the aforementioned cooperative localization algorithms and our proposed FedLoc are the following:
\begin{itemize}
\item In the above mentioned cooperative localization algorithms, the agents (could be a wireless sensor, a robot, or a mobile user) work together to determine their own positions. While in the new proposed FedLoc framework, the mobile users cooperate to train a global, learning-based model that will be used later in the online phase to predict novel positions upon request. 
\item The above mentioned algorithms adopt empirical models, such as the log-distance path-loss model for RSS measurements \cite{PHP03}, and Gaussian mixture model for non-line-of-sight propagation \cite{Yin15b}. In contrast in the FedLoc framework, we solely consider data-driven, machine learning-based models. 
%
\end{itemize}

\subsection{Distributed Target Tracking}
Distributed target tracking is mostly considered for sensor networks without a central node. For such network infrastructures, the traditional Kalman filter or particle filters cannot be used due to lack of the posterior belief/distribution of the desired target state (evolving in time) given all observed sensor measurements. To meet this challenge, various distributed implementations of the Kalman filter and particle {filters}, for instance \cite{Matt02, Gu07, Saber07, Kamgarpour08, Chen11}, were proposed with similar ideas of approximating the posterior belief/distribution as a product of local posteriors. Afterwards, local state estimates are communicated in a message consensus stage. The idea behind these two steps is similar to that of our FedLoc framework. 

However, the major differences between the distributed target tracking and our FedLoc are the following:
\begin{itemize}
\item Distributed Kalman filter and particle filters are based on empirical models, while our FedLoc framework relies on data-driven, machine learning models. 
\item Distributed Kalman and particle filters exchange target state estimates directly over the air, which is fragile to malicious attacks; in contrast FedLoc trains a global deep learning model and advocates changing local model parameters under privacy-preserving schemes. 
\item Distributed Kalman and particle filters do not require training data, but need a good prior distribution of the initial target state. Therefore, they are agile for new deployments. In contrast, our FedLoc framework needs to train the global model before. 
\end{itemize}

\subsection{Crowdsourcing}
Crowdsourcing is a sourcing model in which services are built from a large, relatively open, and often rapidly-evolving group of internet users. Building and maintaining a location system/service based on crowdsourcing is somewhat related to our FedLoc idea. However, the state-of-the-art crowdsourcing methods place more emphasis on raw data sharing and aggregation from a bunch of collaborating users, therefore there is no model in mind. Representative works are as follows. In geography, voluntary users collaboratively build a street map, fill in street information, etc. OpenStreetMap (http://www.openstreetmap.com) and Wikimapia (http:// www.wikimapia.org) are two successful crowdsourcing projects among others. Crowdsourcing of virtual maps, such as RSS map or magnetic map, becomes trendy for big multi-storey buildings \cite{Wu14, Zhang15}. 

The fundamental differences between the crowdsourcing and the FedLoc are the following:
\begin{itemize}
\item Crowdsourcing is more about raw location data aggregation for map construction with less calibration effort, while position determination will be done in a separate stage later on. In contrast, FedLoc focuses on training a global machine learning model for positioning in one step. 
\item Crowdsourcing is mostly model-free. In contrast, FedLoc is built around advanced machine learning models, making it diverse and vibrant.
\item Crowdsourcing aggregates raw data without any safeguard, which will incur severe privacy issues. In contrast, FedLoc processes sensitive data locally and exchanges only the model hyper-parameters that are difficult to decode in general.
\end{itemize}


\subsection{Location Data Modeling}
In this paper, location data specifically refers to spatio-temporal data measured across space as well as time. Representative spatio-temporal data include environmental data, climate data, transportation data, human mobility data, social data, etc. Spatio-temporal data processing and modeling have been well studied over the past decades, ranging from traditional statistical methods to recent data-driven learning model-based methods. Traditional statistical methods include the autoregressive methods for multivariate random fields, factor analysis methods, stochastic process-based methods, tensor decomposition-based methods, see for instance \cite{Diggle13, Atluri2017SpatioTemporalDM}. Data-driven learning models, such as recurrent neural network with long short-term memory and graph neural network have been used to model spatio-temporal data. A comprehensive survey on harnessing deep learning models for spatio-temporal data mining is given in \cite{Wang2019DeepLF}. A special note is given here on the Gaussian process model, which is also called Kriging in geostatistics and can be categorized into the traditional statistical models; however, it can also be regarded as a machine learning model for representing a spatial-temporal function with two inputs, namely the location and the time. In \cite{SSH13,Kuang19}, Gaussian processes implemented via recursive Kalman filtering are used to model spatio-temporal data with rather low computational complexity. Learning models are believed to be able to generate better modeling and prediction performance compared with the traditional statistical methods. In this work, we are keen on training learning models in a distributed manner by a large number of collaborating mobile users. 

\section{Learning Models}
\label{sec:lmodels}
%
This section aims to introduce two representative learning models that can be used as the ``brain'' of the proposed FedLoc framework. We will first briefly review the deep neural network (DNN) model in Subsection~\ref{subsec:dnn}, followed by a short introduction to Gaussian process (GP) model in Subsection~\ref{subsec:gp}. Lastly, we will shed some light on the connections of the two learning models and further highlight the benefits of using GP models over DNN models for FedLoc in Subsection~\ref{subsec:dnnvsGP}. 

\subsection{Deep Neural Network}
\label{subsec:dnn}
Deep neural network (DNN) here refers to the class of feed-forward networks. The term ``feed-forward'' means data are fed from the input layer through several hidden layers to the output layer. Typically, a standard DNN demonstrates a chain structure in math as   
\begin{equation}
	\label{eq:nn_equation}
	f(\boldsymbol{x}; \boldsymbol{\theta}) = f^{(L+1)}(\boldsymbol{W}_{L+1}\cdots f^{(2)}(\boldsymbol{W}_2 f^{(1)}(\boldsymbol{W}_1\boldsymbol{x}))),
\end{equation}
starting from the inputs/features $\boldsymbol{x}$ and passing $L$ hidden layers to the output. In each hidden layer, the mapping function $f^{(i)}(\cdot)$ comprises a bunch of elementary activation functions that mimic the role of neurons in our brain. The commonly used activation functions include the sigmoid function, rectified linear unit (ReLU) function, and some other variants. According to the universal approximation theorem \cite{Hornik91}, a DNN can well approximate any smooth function by tuning the number of hidden layers and the number of neurons in each hidden layer.

One need to train the model hyper-parameters, namely the DNN weights $\boldsymbol{\theta} = \{ \boldsymbol{W}_1, \boldsymbol{W}_2, \dots, \boldsymbol{W}_{L+1} \}$ such that the network output $f(\boldsymbol{x}; \boldsymbol{\theta})$ is close to the  ground truth. Often, DNNs are trained through minimizing the difference between $f(\boldsymbol{x}; \boldsymbol{\theta})$ and $y$. The minimization problem for a set of $n$ training samples can be written as 
\begin{equation}
		\underset{\boldsymbol{\theta}}{\min}~{l}(\boldsymbol{\theta}) := \frac{1}{n}\sum_{i=1}^{n}\ell(y_i, f(\boldsymbol{x}_i; \boldsymbol{\theta})), 
\end{equation} 
where $\ell(\cdot,\cdot)$ is a certain loss function, e.g., the quadratic loss function. Gradient descent type methods with back-propagation are commonly used to solve the above minimization problem in spite of its numerical instability caused by gradient vanishing or explosion.

In the following, we briefly review the batch gradient descent method for DNN training. More details about DNN and gradient descent type method can be found in \cite{Theodoridis15, Goodfellow16}. In general, the iterative training procedure follows three steps:
\begin{itemize}
	\item[1)] Randomly choose a set of weights $\boldsymbol{\theta}^{0}$.
	\item[2)] Iteratively update $\boldsymbol{\theta}^{\eta}$ towards a better $\boldsymbol{\theta}^{\eta+1}$ through
	\begin{equation}
	\boldsymbol{\theta}^{\eta+1} = \boldsymbol{\theta}^{\eta} -\gamma_t \nabla_{\boldsymbol{\theta}}{l}(\boldsymbol{\theta})\vert_{\boldsymbol{\theta} = \boldsymbol{\theta}^{\eta}},
	\end{equation}
	where $\gamma_t$ is the learning rate.
	\item[3)] Repeat step 2) until convergence, e.g., $|{l}(\boldsymbol{\theta}^{\eta}) - {l}(\boldsymbol{\theta}^{\eta+1})| \le \epsilon,$ for some $\epsilon \ge 0$.
\end{itemize}
After obtaining the optimal weights $\hat{\boldsymbol{\theta}}$, one can conduct prediction for a novel input $\boldsymbol{x}_{test}$ using $f(\boldsymbol{x}_{test}; \hat{\boldsymbol{\theta}})$ given in Eq.(\ref{eq:nn_equation}).

\begin{figure}[t]
	\includegraphics[width=0.48\textwidth,height=8cm]{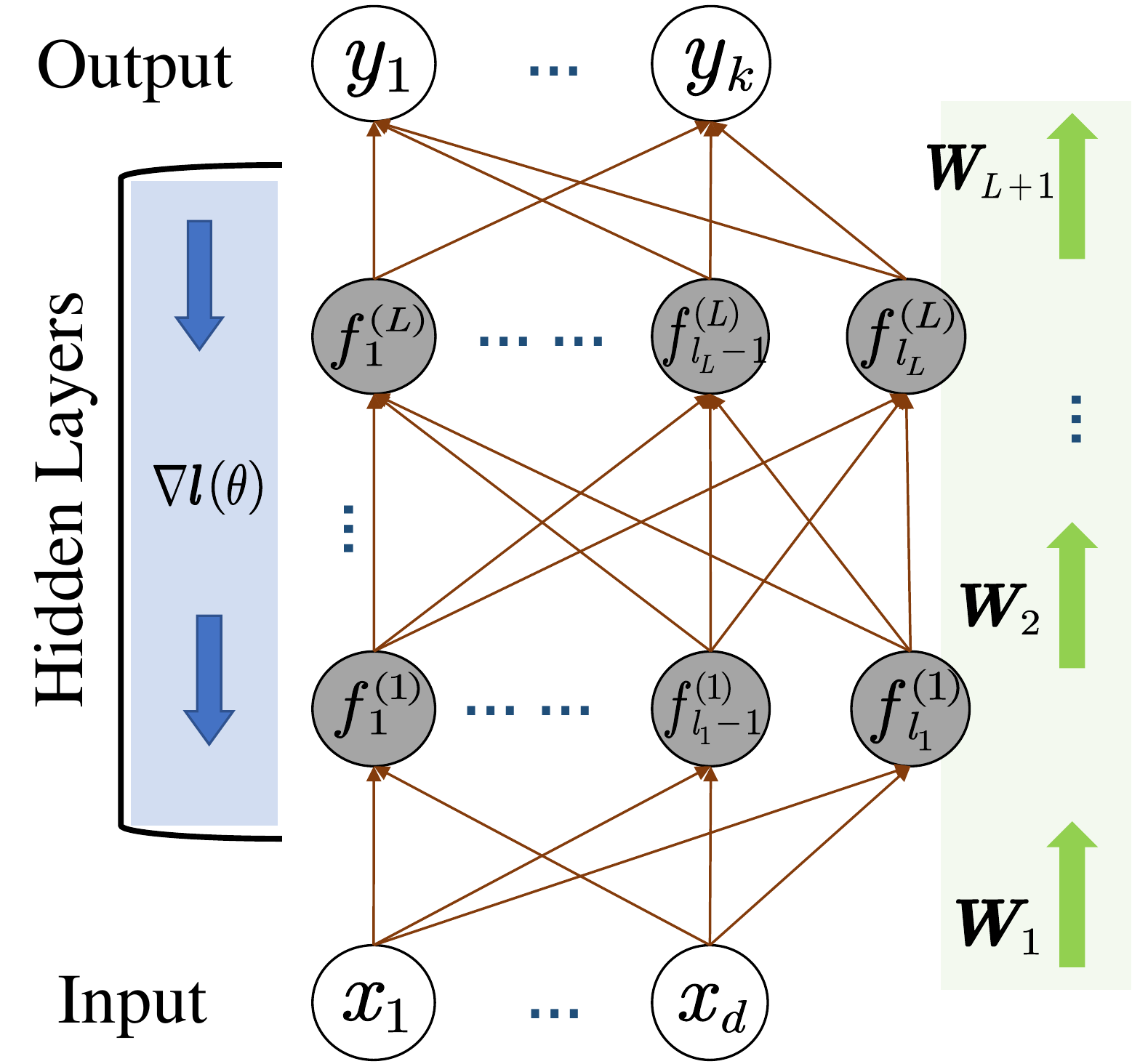}
	\caption{Block diagram of deep neural network architecture. The input-, hidden-, and output variables are represented by nodes, and the weight parameters linking between the nodes at each layer are denoted by $\boldsymbol{W}_i, i\in [1,L+1]$. $\boldsymbol{\theta} = \{ \boldsymbol{W}_1, \boldsymbol{W}_2, \dots, \boldsymbol{W}_{L+1} \}$ comprises all model hyper-parameters, namely the neural network weights of all layers. Green arrows indicate the forward direction of information flow through the network in the inference stage, while the blue arrows indicate the backward direction of the gradient flow for hyper-parameter optimization using back-propagation by default.}
	 \label{fig:neuralnetwork}
\end{figure}

The DNN structure has a big impact on both the forward-propagation and back-propagation computational complexity. For ease of exposition, a specific DNN structure is depicted in Fig.~\ref{fig:neuralnetwork}, wherein we assume $L$ hidden layers and $n$ neurons in each hidden layer, being of the same order as to the number of data samples. Typically, we assume $n \gg L$. Moreover, we assume the number of data samples $n$ is way larger than the feature dimension $d$, i.e., $n \gg d$. For this configuration, the computational complexity required by the forward-propagation is mainly due to the product of the weight matrix and the input vector, namely, $\boldsymbol{W}_{j} f^{(j-1)}(\boldsymbol{x})$, where $j = 1,2,\ldots, L+1$, thus scales as $\mathcal{O}(n^2)$ for one single data sample. The overall computational complexity of the forward-propagation is $\mathcal{O}(n^3)$ for $n$ data samples. As to the back-propagation, let us first note that evaluating $l(\boldsymbol{\theta})$ in each iteration of the gradient descent step requires a forward propagation. Assuming that the gradient descent runs $k$ ($k\ll n$) iterations, the computational complexity for the back-propagation scales as $\mathcal{O}(n^3)$ too.

The aforementioned DNN is suitable for tabular data in general. However, there exist a plethora of deep variants for data with unique features, such as convolutional networks \cite{Lecun10} and capsule networks \cite{Hinton17} for images, long-short-term-memory (LSTM) networks for sequential data, and graph neural networks \cite{kipf2016semisupervised} for spatial and spatio-temporal data. In order to reduce the size of a deep model as well as its computational complexity for use on smartphone and edge devices, one could resort to model distillation techniques \cite{Hinton15} or model sparsification techniques \cite{Frankle19}.

\subsection{Gaussian Processes}
\label{subsec:gp}
Gaussian processes (GP) constitute an important class of Bayesian non-parametric models, which are closely related to several other salient machine learning models. A Gaussian process is a collection of random variables, any finite subset of which follows a Gaussian distribution \cite{RW06}. In the sequel, we solely focus on scalar, real-valued Gaussian processes that are completely specified by a mean function and a kernel function (a.k.a. covariance function). Concretely, 
\begin{equation}
f(\boldsymbol{x}) \sim \mathcal{GP}(m(\boldsymbol{x}), k(\boldsymbol{x}, \boldsymbol{x}'; \boldsymbol{\theta}_{h})),
\label{eq:GP-model} 
\end{equation} 
where $m(\boldsymbol{x}) $ is the mean function, which is often set to zero in practice, especially when there is no prior knowledge about the underlying process; and $k(\boldsymbol{x}, \boldsymbol{x}'; \boldsymbol{\theta}_{h})$ is the kernel function tuned by the kernel hyper-parameters, $\boldsymbol{\theta}_{h}$.

Let us consider the GP regression model, $y = f(\boldsymbol{x}) + e$,
%
where $y \in \mathbb{R}$ is a continuous-valued, scalar output; the unknown function $f(\boldsymbol{x}) : \mathbb{R}^{d} \mapsto \mathbb{R}$ is modeled as a zero mean GP; and the noise $e$ is assumed to be Gaussian distributed with zero mean and variance $\sigma_{e}^{2}$. Moreover, the noise terms at different data points are assumed to be mutually independent. The set of all unknown GP hyper-parameters is denoted by $\boldsymbol{\theta} \triangleq [\boldsymbol{\theta}_{h}^{T}, \sigma^{2}_{e}]^T$, and the dimension of $\boldsymbol{\theta}$ is assumed to be equal to $p$.

%
The joint prior distribution of the training output $\boldsymbol{y}$ and test output $\boldsymbol{y}_{*}$ can be written compactly as
\begin{equation}
\begin{bmatrix} \boldsymbol{y} \\ \boldsymbol{y}_{*} \end{bmatrix} \sim \mathcal{N} \left( \boldsymbol{0},  \begin{bmatrix} \boldsymbol{K}(\boldsymbol{X}, \boldsymbol{X}) + \sigma_{e}^{2} \boldsymbol{I}_{n}, & \!\!\!\!\!\!\!\! \boldsymbol{K}(\boldsymbol{X}, \boldsymbol{X}_{*}) \\ \boldsymbol{K}(\boldsymbol{X}_{*}, \boldsymbol{X}), & \!\!\!\!\!\!\!\! \boldsymbol{K}(\boldsymbol{X}_{*}, \boldsymbol{X}_{*}) + \sigma_{e}^{2} \boldsymbol{I}_{n_{*}} \end{bmatrix} \right),  
\end{equation}
where $\boldsymbol{K}(\boldsymbol{X}, \boldsymbol{X})$ is an $n \times n$ covariance matrix between the training inputs; $\boldsymbol{K}(\boldsymbol{X}, \boldsymbol{X}_{*})$ is an $n \times n_{*}$ covariance matrix between the training inputs and test inputs, $\boldsymbol{K}(\boldsymbol{X}_{*}, \boldsymbol{X}_{*})$ is an $n_{*} \times n_{*}$ covariance matrix between the test inputs. Here, we let $\boldsymbol{K}(\boldsymbol{X}, \boldsymbol{X})$ be the short term of $\boldsymbol{K}(\boldsymbol{X}, \boldsymbol{X}; \boldsymbol{\theta}_h)$.

Applying some known results of conditional Gaussian distribution, we can easily derive the posterior distribution as
\begin{equation}
p(\boldsymbol{y}_{*} \vert \mathcal{D}, \boldsymbol{X}_{*}; \boldsymbol{\theta}_h) \sim
\mathcal{N} \left(  \bar{\boldsymbol{m}} , \bar{\boldsymbol{V}}  \right), 
\end{equation} 
where the posterior mean (vector) and the posterior covariance (matrix) are respectively,
\begin{align}
\bar{\boldsymbol{m}} & = \boldsymbol{K}(\boldsymbol{X}_{*}, \boldsymbol{X}) \left[ \boldsymbol{K}(\boldsymbol{X}, \boldsymbol{X}) + \sigma_{e}^{2} \boldsymbol{I}_{n} \right]^{-1} \boldsymbol{y}, \\
\bar{\boldsymbol{V}} & = \boldsymbol{K}(\boldsymbol{X}_{*}, \boldsymbol{X}_{*}) + \sigma_{e}^{2} \boldsymbol{I}_{n_{*}} \nonumber \\ 
&- \boldsymbol{K}(\boldsymbol{X}_{*}, \boldsymbol{X}) \left[ \boldsymbol{K}(\boldsymbol{X}, \boldsymbol{X}) + \sigma_{e}^{2} \boldsymbol{I}_{n} \right]^{-1}  \boldsymbol{K}(\boldsymbol{X}, \boldsymbol{X}_{*}).
\end{align}
Given a novel input in the test data set, the above posterior mean gives the prediction, while the posterior covariance gives the uncertainty region of the prediction.
\begin{figure}[t]
	\centering
	\subfloat[Prior]{\includegraphics[width=8.2cm]{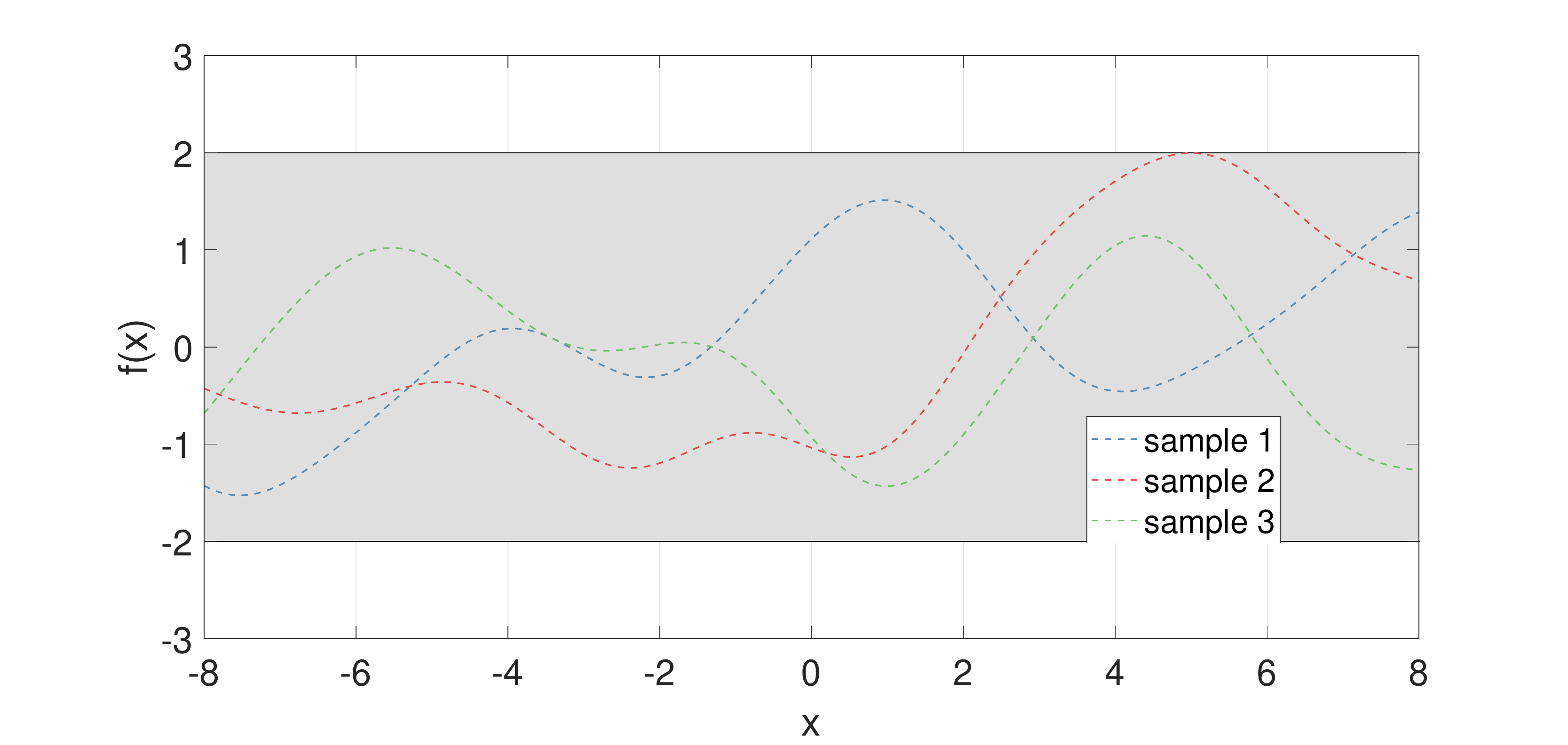}}
	
	\subfloat[Posterior]{\includegraphics[width=8cm]{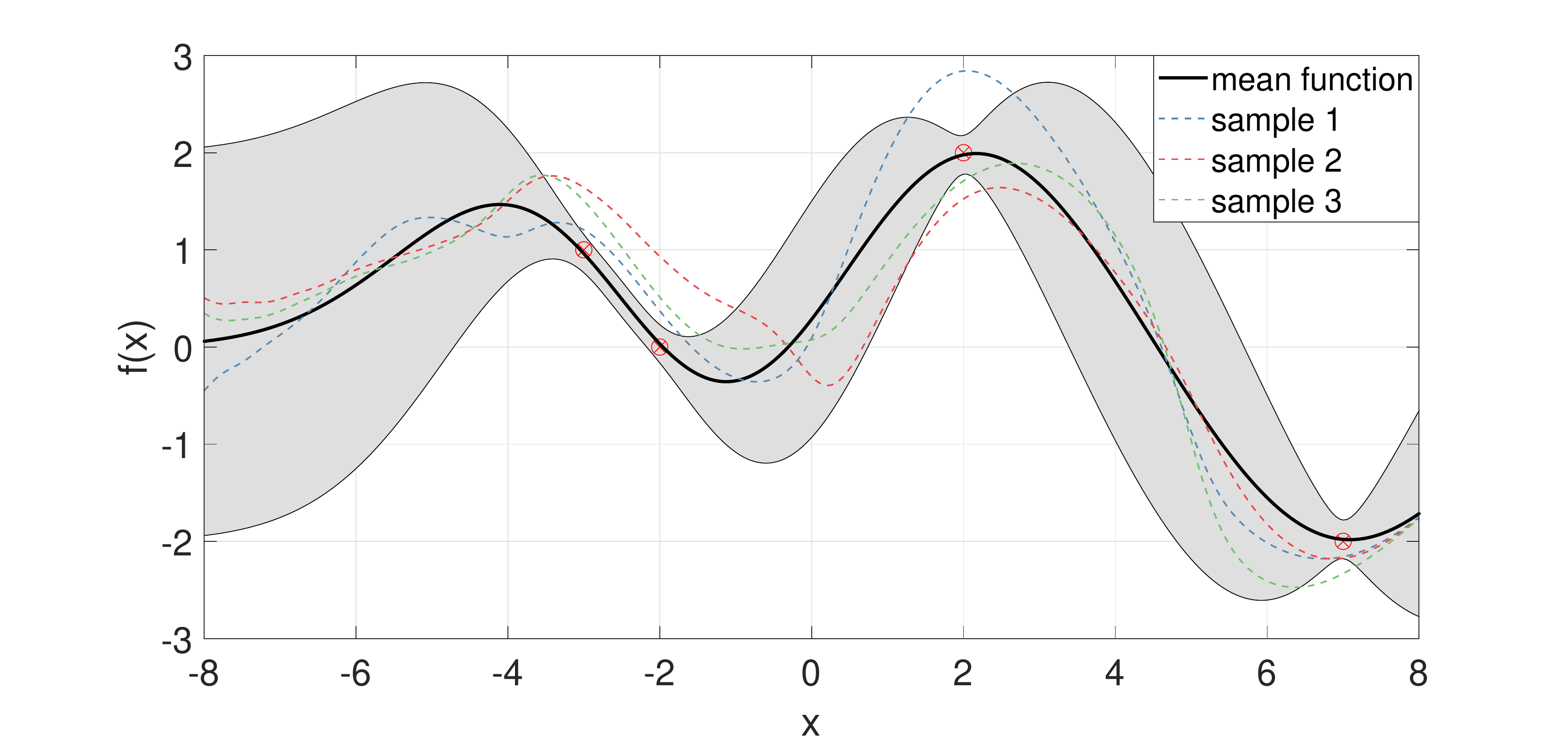}}
	\caption{Subfigure (a) shows three sample functions drawn randomly from a GP prior with a specific squared-exponential kernel. Subfigure (b) shows three sample functions drawn from the posterior conditioned on the prior in (a) as well as four noisy observations indicated by red dots. The corresponding posterior mean function is depicted by the black curve. The grey shaded area represents the uncertainty region, namely the $95\%$ confidence region for both the prior and the posterior, respectively.}
	\label{fig:gp_mode}
\end{figure}

Kernel function determines the power of the GP model to a large extent. In order to make a kernel function full of expressive power and automatically adaptive to a given data set, the following works can be adopted. In \cite{WA13}, a spectral mixture (SM) kernel was proposed to approximate the spectral density with a Gaussian mixture model arbitrarily well in the frequency domain and transform it back into a universal stationary kernel. In \cite{Yin18}, the authors modified the SM kernel to a linear multiple low-rank sub-kernels with a favorable optimization structure, which enables faster and more stable numerical search. In \cite{Wilson16, wilson2016stochastic, al2017learning, Xue19}, a DNN architecture was combined with the automatic relevance determination (ARD) kernel to approximate any kernel function (including both the stationary and non-stationary ones). Yet, in a more recent trend designing universal kernels may be obtained as a byproduct of designing new fashioned deep GP models \cite{Damianou13} that link DNNs to GPs \cite{Cho09, Mattews18, Lee18}.

Next, we introduce the classical ML-based GP hyper-parameter estimation. Due to the Gaussian assumption on the noise, the log-likelihood function can be obtained in closed form. The GP hyper-parameters can be optimized equivalently by minimizing the negative log-likelihood function:
\begin{equation}
 l(\boldsymbol{X}, \boldsymbol{y}; \boldsymbol{\theta}) = \boldsymbol{y}^T \boldsymbol{C}^{-1}(\boldsymbol{\theta}) \boldsymbol{y} +  \log \det \left( \boldsymbol{C}(\boldsymbol{\theta}) \right), 
\label{eq:likelihood}
\end{equation}
where $\boldsymbol{C}(\boldsymbol{\theta}) \triangleq \boldsymbol{K}(\boldsymbol{X}, \boldsymbol{X}; \boldsymbol{\theta}_h) + \sigma_{e}^{2} \boldsymbol{I}_{n}$. This optimization problem is mostly solved via gradient descent type methods, such as LFGS-Newton or conjugate gradient \cite{RW06}, which requires the following closed-form partial derivatives, for $i=1,2,...,p$,
\begin{equation}
\frac{\partial l(\boldsymbol{\theta}) }{\partial \theta_i} \!=\!  tr \!\left( \boldsymbol{C}^{-1}(\boldsymbol{\theta})  \frac{\partial \boldsymbol{C}(\boldsymbol{\theta})}{\partial \theta_i} \right)  -  \boldsymbol{y}^{T} \boldsymbol{C}^{-1}(\boldsymbol{\theta})  \frac{\partial \boldsymbol{C}(\boldsymbol{\theta})}{\partial \theta_i} \boldsymbol{C}^{-1}(\boldsymbol{\theta}) \boldsymbol{y}.  
\end{equation}

It should be noted that the minimization problem in Eq.(\ref{eq:likelihood}) may easily get stuck at a bad local optimum when the selected learning model is over-parameterized and the associated cost function does not show any favorable optimization structure.

Using the above ML method to train a GP model requires $O(n^3)$ computational complexity thus forbids its practical use. To address this difficulty, a plethora of scalable GP models have been developed in the past decades. Some representative works of different categories were obtained through using (1) low-rank kernel matrix approximation \cite{williams2001}; (2) local structures of the kernel matrix \cite{Ambikasaran16}; (3) the state-space model reformulation and Kalman filter \cite{SSH13}; (4) the Bayesian committee machine (BCM) with a number of distributed computing units \cite{Deisenroth15}; and (5) the variational Bayesian formulation \cite{Titsias09}. A comprehensive survey of the existing scalable GP models can be found in \cite{liu2020gaussian}.

\subsection{DNN Versus GP}
\label{subsec:dnnvsGP}
In the previous subsections, we briefly introduced DNN and GP that can both be used as the core learning model. DNN is quite popular nowadays due to various good reasons. Among others, it can approximate any smooth function according to the universal approximation theorem \cite{Hornik91}. But the main drawbacks of DNN lie in its opaque model interpretability and the large number of hyper-parameters (DNN weights) to be trained. For our FedLoc framework proposed in this paper, we put more emphasis on the GP models due to their unique welcome features compared with DNN. 

First, GP models involve significantly fewer model hyper-parameters than an equally-effective DNN. From \cite{Neal95} we know that a single layer Bayesian neural network with i.i.d. weights converges to a GP. Consequently, a neural network kernel was designed with the following explicit form \cite{RW06}:
\begin{equation}
k_{\text{NN}}(\boldsymbol{x}, \boldsymbol{x}') = \frac{2}{\pi} \sin^{-1} \left( \frac{2 \tilde{\boldsymbol{x}} \Sigma \tilde{\boldsymbol{x}}' }{ \sqrt{(1+ 2 \tilde{\boldsymbol{x}} \Sigma \tilde{\boldsymbol{x}}) (1+ 2 \tilde{\boldsymbol{x}}' \Sigma \tilde{\boldsymbol{x}}')} }  \right) ,
\end{equation}
where $\tilde{\boldsymbol{x}} \triangleq [1, \boldsymbol{x}^T]^T$ is an augmented input vector. Often, we assume $\Sigma = diag(\sigma_{1}^2, \sigma_{2}^2,...,\sigma_{d+1}^2)$ to be a diagonal matrix, thus the hyper-parameters $\boldsymbol{\theta}_{h} = [\sigma_{1}^2, \sigma_{2}^2,...,\sigma_{d+1}^2]^{T}$ is of dimension $d+1$. If $\Sigma$ is taken to be a general matrix, the hyper-parameters to be tuned is in the order of $d^2$, being much smaller than the size of a fully-connected DNN in general.

Lately, the arc-cosine kernel \cite{Cho09}, the neural tangent kernel (NTK) \cite{Jacot18}, and the convolutional neural tangent kernel (CNTK) \cite{Arora2019} were developed to mimic a DNN with infinite width. The arc-cosine kernel function \cite{Cho09} is given by
\begin{equation}
	\label{eq:arccosine}
	\begin{aligned}
		&k_{\text{arccos}}(\boldsymbol{x}, \boldsymbol{x^\prime}) = \\
		& ~~~2 \int  \frac{e^{-\frac{\left\lVert \boldsymbol{\mathrm{w}}\right\rVert ^ 2}{2}}}{(2\pi)^{d/2}} \Theta (\boldsymbol{\mathrm{w}} \cdot \boldsymbol{x})  \Theta (\boldsymbol{\mathrm{w}} \cdot \boldsymbol{x^\prime} ) (\boldsymbol{\mathrm{w}} \cdot \boldsymbol{x} ) ^ q (\boldsymbol{\mathrm{w}} \cdot \boldsymbol{x^\prime} ) ^ q \mathrm{d} \boldsymbol{\mathrm{w}},
	\end{aligned}
\end{equation}
where $\Theta(z) = \frac{1}{2} (1 + \operatorname{sign}(z))$ denotes the Heaviside step function, and $q$ is a non-negative integer for selecting a particular activation function. The arc-cosine kernel for multi-layer neural network can also be obtained via a  recursive kernel design. The hyper-parameters of the arc-cosine kernel include the kernel order parameter $q$ for specifing the activation function and the number of hidden layers $L$.

The NTK captures the behavior of fully-connected deep neural networks trained by gradient descent, and CNTK is an extension of NTK to convolutional neural networks. 
The analytic form of NTK can be derived recursively as
\begin{equation}
	\label{eq:NTK}
    k_{\text{NTK}}\left(\boldsymbol{x}, \boldsymbol{x}^{\prime}\right)=\sum_{h=1}^{L+1}\left(\Sigma^{(h-1)}\left(\boldsymbol{x}, \boldsymbol{x}^{\prime}\right) \cdot \prod_{h^{\prime}=h}^{L+1} \dot{\Sigma}^{\left(h^{\prime}\right)}\left(\boldsymbol{x}, \boldsymbol{x}^{\prime}\right)\right),
  \end{equation}
where $\Sigma^{(h-1)}$ is the centered covariance matrix of the $(h-1)$th layer's output $f^{(h)}(\boldsymbol{x})$, and $\dot{\Sigma}$ is the corresponding derivative covariance. 

It can be proven that a sufficiently wide and randomly initialized DNN trained by gradient descent is equivalent to a kernel regression predictor with the aforementioned NTK kernel. Hence, the properties of the ultra-wide DNN, such as the generalization capability, can be obtained by learning the corresponding NTK, albeit with {much less computational} effort. It is also noteworthy that the hyper-parameter of the NTK is only the number of layers that can be tuned easily using cross-validation.


Second, GP models can handle input uncertainty naturally. In our considered applications, the model inputs often involve position or position related measures that are intrinsically subject to noise due to imperfect field calibration. Since GP model is a probabilistic model, the input uncertainty can be easily incorporated into the model. One way is to assume the training input $\boldsymbol{x}$ to be a random variable with a known distribution $p(\boldsymbol{x})$. In \cite{Girard04}, for instance, the mean function of GP with input uncertainty was obtained as
\begin{equation}
\tilde{m}(\boldsymbol{x}) = \int m(\boldsymbol{x}) p(\boldsymbol{x}) \textit{d} \boldsymbol{x},
\end{equation}       
and the kernel function obtained as
\begin{align}
\tilde{k}(\boldsymbol{x}, \boldsymbol{x}') = \int \!\!\! \int  k(\boldsymbol{x}, \boldsymbol{x}') p(\boldsymbol{x}) p(\boldsymbol{x}') \textit{d} \boldsymbol{x} \textit{d} \boldsymbol{x}'. 
\end{align}
The only difficulty lies in the evaluation of the two integrals. In general, they can be approximated by Monte-Carlo integration \cite{Theodoridis15,Bishop06}. The rest of the steps remain the same as the standard GP with clean input as given in (1). The computation can be largely reduced for Gaussian distributed input $\boldsymbol{x}$ using unscented transform, see for instance \cite[Chapter~5.5]{Sarkka13}. 

Third, GP models can more easily encode prior information about the data than DNN. This is  inherited from the meaningful interpretation of various elementary kernels with known characteristics. For instance, when the data demonstrate periodicity, we could add elementary periodic kernel(s) or locally periodic kernel(s) to the eventual composite kernel; when the data demonstrate linear rising trend, we could add a linear kernel to the eventual kernel; when the data profile is verified to be smooth, we could use the squared-exponential (SE) kernel with a large length scale parameter. Taking into account the prior information about the data can be regarded as regularizing the model fitting process, thus is effective for avoiding data over-fitting. This is a welcome feature for our applications in which the total amount of data is large but each mobile user may only have a small amount of local data in hand for training the global model. According to a recent white paper released by Huawei, wireless big data in 6G will be generated by a huge amount of mobile users and IoT devices, each contributing only a small local data set. 

Finally, it is noteworthy that DNNs and their variants are still more widely used than GPs for machine learning empowered applications. But for localization applications, yet, GP models are very promising due to the aforementioned advantages. 

\section{Federated Localization (FEDLOC)}
\label{sec:fwp}
The organization of this section is the following. In Subsection~\ref{subsec:FLreview}, the main idea of federated learning is introduced, followed by a review of various existing distributed training methods proposed for DNN and GP learning models in Subsection~\ref{subsec:distOpt}. Privacy-preserving schemes are briefly surveyed in Subsection~\ref{subsec:privacy}. Lastly, we conclude this section by giving a full picture of the FedLoc framework.

\subsection{Brief Review of Federated Learning}
\label{subsec:FLreview}
The idea of federated learning exists for a long time in the context of distributed learning, and it was given the name by some researchers at Google in 2016 \cite{Konecny16, mcmahan17a}. Federated learning is a flexible and safe cooperation framework for mobile users. The idea behind the federated learning is to approximate a global model/objective as a summation of local models/objectives trained individually by mobile users. Mathematically, the above idea can be expressed as 
\begin{equation}
l(\boldsymbol{X}, \boldsymbol{y}; \boldsymbol{\theta}) \approx \sum_{k=1}^{K} l^{(k)}(\boldsymbol{X}_k, \boldsymbol{y}_k; \boldsymbol{\theta}), 
\label{eq:local-approx}
\end{equation}
where $\boldsymbol{X}$ is the complete set of the training inputs, $\boldsymbol{y}$ is the complete set of the training outputs, and they constitute the complete training set $\mathcal{D}$; $l(\cdot)$ is a global objective in terms of the model hyper-parameters $\boldsymbol{\theta}$; while $\boldsymbol{X}_k$ is the $k$-th local set of the training inputs, $\boldsymbol{y}_k$ is the $k$-th local set of the training outputs, and they constitute $\mathcal{D}_k$, which is a subset of $\mathcal{D}$; $l^{(k)}(\cdot)$ is a local objective of the $k$-th local data set, $\mathcal{D}_k$; $K$ is the total number of collaborating mobile users, which is assumed to be large. Both $l(\cdot)$ and $l^{(k)}(\cdot)$ are composite functions of a selected learning model/regression function and a cost function. Lastly, we note that the outputs $\boldsymbol{y}$ are mostly positions or position related measurements in our work. 

To shed some light on the objective $l(\cdot)$, let us consider the following two different machine learning models and their cost functions.

\textbf{I: DNN model with the Least-Squares Cost.} 
The global objective for training a DNN is given as follows:
\begin{equation}
l(\boldsymbol{X}, \boldsymbol{y}; \boldsymbol{\theta}) = \sum_{i=1}^{n} \left( y_i - f(\boldsymbol{x}_i; \boldsymbol{\theta}) \right)^2,
\end{equation}
where the outputs are assumed to be independent, and $f(\boldsymbol{x}_i; \boldsymbol{\theta})$ is represented by a DNN with $L$ hidden layers \cite{Goodfellow16} with $\boldsymbol{\theta} = \{ \boldsymbol{W}_1, \boldsymbol{W}_2,..., \boldsymbol{W}_{L+1} \}$ representing the DNN weights to be tuned for all hidden layers. It is obvious that the global objective is already in form of sum-of-residual-squared.

\textbf{II: GP model with the Maximum Likelihood Cost.} 
Due to the Gaussian assumption on the noise, the log-likelihood function can be obtained in closed form. Therefore, the global objective for training the GP regression model hyper-parameters is
\begin{align}
l(\boldsymbol{X}, \boldsymbol{y}; \boldsymbol{\theta}) &= \log p(\boldsymbol{y}; \boldsymbol{X},\boldsymbol{\theta}) \nonumber \\
&= \log \mathcal{N} \left( \boldsymbol{y}; \boldsymbol{m}(\boldsymbol{X}), K(\boldsymbol{X}, \boldsymbol{X}; \boldsymbol{\theta})\right),
\label{eq:GP-global-obj}
\end{align}
where the vector $\boldsymbol{m}(\boldsymbol{X})$ and the matrix $K(\boldsymbol{X}, \boldsymbol{X}; \boldsymbol{\theta})$ are respectively the mean function $m(\boldsymbol{x})$ and the kernel function $k(\boldsymbol{x}, \boldsymbol{x}'; \boldsymbol{\theta})$ evaluated for the complete data set $\mathcal{D}$. This global objective is not directly in the form of summation, but commonly approximated by the product-of-expert (PoE) \cite{Deisenroth15} as 
\begin{equation}
l(\boldsymbol{X}, \boldsymbol{y}; \boldsymbol{\theta}) \approx \sum_{i=1}^{K} \log \mathcal{N} \left( \boldsymbol{y}_k; \boldsymbol{m}(\boldsymbol{X}_k), K(\boldsymbol{X}_k, \boldsymbol{X}_k; \boldsymbol{\theta}) \right).
\label{eq:GP-local-approx}
\end{equation}
Here, we note that the independent noise term has been absorbed into the kernel function for notation brevity in Eq.(\ref{eq:GP-global-obj}) and Eq.(\ref{eq:GP-local-approx}).

\subsection{Distributed Training of the Learning Models}
\label{subsec:distOpt}
The original goal is to train a global model through
\begin{equation}
\hat{\boldsymbol{\theta}} = \arg \min_{\boldsymbol{\theta}} l(\boldsymbol{X}, \boldsymbol{y}; \boldsymbol{\theta}), 
\end{equation}
where the objective function is often non-convex and solved by gradient descent type methods. 
%
When the complete data set is large, training the global model given above can be computationally expensive. As mentioned before, federated learning aims to distribute the heavy computation load to a massive number of collaborating mobile users by solving instead the following problem:
\begin{equation}
\hat{\boldsymbol{\theta}} = \arg \min_{\boldsymbol{\theta}} \sum_{k=1}^{K} l^{(k)}(\boldsymbol{X}_k, \boldsymbol{y}_k; \boldsymbol{\theta}).
\label{eq:dist-approx}
\end{equation}
Each mobile user maintains a local update of the global model hyper-parameters and sends it to a central node for consensus. There exist various ways for updating the global model hyper-parameters. In the following, we introduce the classical federated averaging (FedAvg) \cite{mcmahan17a} algorithm and a few algorithms developed upon alternating direction of multipliers method (ADMM) \cite{Boyd11, Hong16}. 

\begin{figure}
	\centering
	\footnotesize
	\subfloat[]{
		\begin{tikzpicture}[level 1/.style={sibling distance=0.31\columnwidth}, align=center, every node/.style={rounded corners}]
		\node[draw, text width=0.65\columnwidth, text height=0.7em](C){global gradient descent step for $\boldsymbol{\theta}^{r+1}$}
		child{
			node[draw](S1){$\nabla l^{(1)}(\boldsymbol{\theta}^{r+1})$}
			edge from parent[draw=none]
		}
		child{
			node[draw](S2){$\nabla l^{(2)}(\boldsymbol{\theta}^{r+1})$}
			edge from parent[draw=none]
		}
		child{
			node[draw](S3){$\nabla l^{(K)}(\boldsymbol{\theta}^{r+1})$}
			edge from parent[draw=none]
		};
		\draw[-latex] (S1.80) -- (S1.80|-C.south) node[midway, right, align=left] {$\nabla l^{(1)}(\boldsymbol{\theta}^r)$};
		\draw[-latex] (S1.100|-C.south) -- (S1.100) node[midway, left, text width=0.11\columnwidth, align=right] {$\boldsymbol{\theta}^{r+1}$};
		\draw[-latex] (S2.80) -- (S2.80|-C.south) node[midway, right, align=left] {$\nabla l^{(2)}(\boldsymbol{\theta}^r)$};
		\draw[-latex] (S2.100|-C.south) -- (S2.100) node[midway, left, align=right] {$\boldsymbol{\theta}^{r+1}$};
		\draw[-latex] (S3.80) -- (S3.80|-C.south) node[midway, right, align=left] {$\nabla l^{(K)}(\boldsymbol{\theta}^r)$};
		\draw[-latex] (S3.100|-C.south) -- (S3.100) node[midway, left, align=right] {$\boldsymbol{\theta}^{r+1}$};
		\path (S2) -- (S3) node [midway] {$\cdots$};
		\end{tikzpicture}
		\label{subfig:gradient descent}}
	
	\subfloat[]{
		\begin{tikzpicture}[level 1/.style={sibling distance=0.31\columnwidth}, align=center, every node/.style={rounded corners}]
		\node[draw, text width=0.65\columnwidth, text height=0.7em](C){global consensus for $\boldsymbol{z}^{r+1}$}
		child{
			node[draw](S1){$\boldsymbol{\theta}_{1}^{r+1}, \boldsymbol{\beta}_{1}^{r+1}$}
			edge from parent[draw=none]
		}
		child{
			node[draw](S2){$\boldsymbol{\theta}_{2}^{r+1}, \boldsymbol{\beta}_{2}^{r+1}$}
			edge from parent[draw=none]
		}
		child{
			node[draw](S3){$\boldsymbol{\theta}_{K}^{r+1}, \boldsymbol{\beta}_{K}^{r+1}$}
			edge from parent[draw=none]
		};
		\draw[-latex] (S1.80) -- (S1.80|-C.south) node[midway, right, align=left] {$\boldsymbol{\theta}_{1}^{r} + \frac{1}{\rho_1}\boldsymbol{\beta}_{1}^{r}$};
		\draw[-latex] (S1.100|-C.south) -- (S1.100) node[midway, left, text width= 0.11\columnwidth, align=right] {$\boldsymbol{z}^{r+1}$};
		\draw[-latex] (S2.80) -- (S2.80|-C.south) node[midway, right, align=left] {$\boldsymbol{\theta}_{2}^{r} + \frac{1}{\rho_2}\boldsymbol{\beta}_{2}^{r}$};
		\draw[-latex] (S2.100|-C.south) -- (S2.100) node[midway, left, align=right] {$\boldsymbol{z}^{r+1}$};
		\draw[-latex] (S3.80) -- (S3.80|-C.south) node[midway, right, align=left] {$\boldsymbol{\theta}_{K}^{r} + \frac{1}{\rho_K}\boldsymbol{\beta}_{K}^{r}$};
		\draw[-latex] (S3.100|-C.south) -- (S3.100) node[midway, left, align=right] {$\boldsymbol{z}^{r+1}$};
		\path (S2) -- (S3) node [midway] {$\cdots$};
		\end{tikzpicture}
		\label{subfig:ADMM}}
	\caption{Workflow of two existing distributed hyper-parameter optimization schemes. (a) FedAvg \cite{mcmahan17a}. (b) cADMM \cite{Boyd11}. }
\end{figure}
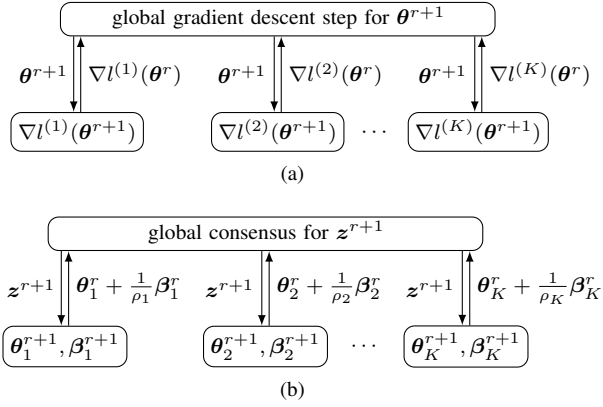

We start with the state-of-the-art FedAvg algorithm. Typically, the $k$-th mobile user calculates the gradient $\nabla l^{(k)}(\boldsymbol{\theta})$ and uploads it to the central node. The central node then aggregates a batch of/all local gradients to approximate $\nabla_{\boldsymbol{\theta}} l(\boldsymbol{X}, \boldsymbol{y}; \boldsymbol{\theta})$. We illustrate this workflow in Fig.~\subref*{subfig:gradient descent}, which is named by FedAvg and deemed as the optimization algorithmic core of the federated learning framework \cite{mcmahan17a}. A robust variant, called FedProx \cite{Sahu2018}, was proposed to improve local training convergence by adding an extra proximal step at each client to restrict the distance between the local parameter estimates and the current global estimate.

Next, we introduce two ADMM-based hyper-parameter optimization schemes, that can effectively balance the computation and communication efficiency. The first one, namely the classical ADMM-based hyper-parameter optimization scheme (short as cADMM), reformulates the optimization problem in (\ref{eq:dist-approx}) as a nonconvex consensus problem \cite{Boyd11} with a set of newly introduced local hyper-parameters $ \{ \boldsymbol{\theta}_1, \boldsymbol{\theta}_2, \ldots, \boldsymbol{\theta}_K \} $ and the global hyper-parameter $\boldsymbol{z}$. Concretely, we solve instead
\begin{equation}\label{P2}
\begin{aligned}
& \min \quad \textstyle{\sum_{k=1}^{K}} l^{(k)}(\boldsymbol{\theta}_k), \\
& \,\,\, \mathrm{s.t.}  \quad \boldsymbol{\theta}_k - \boldsymbol{z} = \boldsymbol{0}, \quad \forall \, k=1,2, \ldots, K,
\end{aligned}
\end{equation}
%
%
where $l^{(k)}(\boldsymbol{\theta}_k)$ is nonconvex in terms of the local hyper-parameter $\boldsymbol{\theta}_k$ in general. The augmented Lagrangian function for Eq.(\ref{P2}) is given by
\begin{eqnarray}
\mathcal{L}(\{\boldsymbol{\theta}_{k}\}, \boldsymbol{z}, \{\boldsymbol{\beta}_{k}\}) \!\! &=& \!\! \textstyle{\sum_{k=1}^{K}} (l^{(k)}(\boldsymbol{\theta}_{k}) + \boldsymbol{\beta}_{k}^{T}(\boldsymbol{\theta}_{k} - \boldsymbol{z}) \nonumber\\
&& \negmedspace {} + (\rho_k/2)\|\boldsymbol{\theta}_{k} - \boldsymbol{z}\|_{2}^{2}),
\end{eqnarray}
where $\boldsymbol{\beta}_k$ is a dual variable, and $\rho_k$ stands for a predetermined regularization parameter. The $(r+1)$-th iteration of the cADMM for solving (Eq.\ref{P2}) can be decomposed into
   \begin{subequations}
	\begin{eqnarray}
	\boldsymbol{z}^{r+1} & = & \frac{1}{K}\textstyle{\sum_{k=1}^{K}} ( \boldsymbol{\theta}_{k}^{r} + \frac{1}{\rho_k}\boldsymbol{\beta}_{k}^{r} ) ,\\
	\boldsymbol{\theta}_{k}^{r+1} & = & \textstyle{\argmin_{\boldsymbol{\theta}_{k}}}(l^{(k)}(\boldsymbol{\theta}_{k}) +  (\boldsymbol{\beta}_{k}^{r})^{T}(\boldsymbol{\theta}_{k} - \boldsymbol{z}^{r+1}) \nonumber \\
	&& \negmedspace {} + (\frac{\rho_i}{2})\|\boldsymbol{\theta}_{k} - \boldsymbol{z}^{r+1}\|_{2}^{2}), \label{eq:theta minimization}\\
	\boldsymbol{\beta}_{k}^{r+1} & = & \boldsymbol{\beta}_{k}^{r} + \rho_k(\boldsymbol{\theta}_{k}^{r+1} - \boldsymbol{z}^{r+1}).
	\end{eqnarray}
	\end{subequations}
The above workflow is shown in Fig.~\subref*{subfig:ADMM} for clarity.

Next, we continue to introduce a more recent proximal ADMM (short as pxADMM) scheme proposed in \cite{Hong16}, which is capable of reducing the communication overhead and the computational time \emph{at the same time}. Unlike in step Eq.\eqref{eq:theta minimization} where the local hyper-parameters $\boldsymbol{\theta}_{k}$ are updated through minimizing the augmented Lagrangian function exactly, the proximal ADMM takes a proximal step w.r.t.\ $\boldsymbol{\theta}_{k}$ by applying the first-order Taylor expansion to $l^{(k)}(\boldsymbol{\theta}_{k})$ \cite{Hong16}, i.e.,
\begin{eqnarray}\label{eq:proximal step}
\boldsymbol{\theta}_{k}^{r+1} \!\!\!\!\! & = & \!\!\!\!\! \textstyle{\argmin_{\boldsymbol{\theta}_{k}}}  \nabla^{T}l^{(k)}(\boldsymbol{z}^{r+1})(\boldsymbol{\theta}_{k} - \boldsymbol{z}^{r+1})  \nonumber \\ 
& + & \!\!\!\!\!\! (\boldsymbol{\beta}_{k}^{r})^{T}(\boldsymbol{\theta}_{k} \!\!-\!\! \boldsymbol{z}^{r+1}) \!\!+\!\! \left(\frac{\rho_k \!+\! L_k}{2}\right) \!\! \|\boldsymbol{\theta}_{k} \!-\! \boldsymbol{z}^{r+1}\|_{2}^{2} , 
\end{eqnarray}
where $L_k$ is a newly introduced positive constant making $\| \nabla l^{(k)}(\boldsymbol{\theta}_{k}) - \nabla l^{(k)}(\boldsymbol{\theta}_{k}') \| \leq L_k \| \boldsymbol{\theta}_{k} - \boldsymbol{\theta}_{k}' \|$ satisfied for all $\boldsymbol{\theta}_{k} \text{ and } \boldsymbol{\theta}_{k}',  k=1,2, \ldots, K$. Note that the proximal step in Eq.\eqref{eq:proximal step} for $\boldsymbol{\theta}_k$ is a (convex) quadratic optimization problem with the following closed-form solution:
%
%
%
\begin{eqnarray}\label{eq:theta update}
\boldsymbol{\theta}_{k}^{r+1} 
\!\! & = & \!\! \boldsymbol{z}^{r+1} - \left(\frac{\nabla l^{(k)}(\boldsymbol{z}^{r+1}) + \boldsymbol{\beta}_k^r}{\rho_k + L_k} \right).
\end{eqnarray}
As a consequence, the $(r+1)$-th iteration of the pxADMM for solving Eq.(\ref{P2}) can be decomposed into
\begin{subequations}
\begin{eqnarray}
%
\boldsymbol{z}^{r+1} \!\!\! & = & \!\!\! (1/K)\textstyle{\sum_{k=1}^{K}} ( \boldsymbol{\theta}_{k}^{r} + \frac{1}{\rho_i}\boldsymbol{\beta}_{k}^{r} ), \\
\boldsymbol{\theta}_k^{r+1} \!\!\! & = & \!\!\! \boldsymbol{z}^{r+1} - \frac{(\nabla l^{(k)}(\boldsymbol{z}^{r+1}) + \boldsymbol{\beta}_k^r)}{\rho_k + L_k}, \label{eq:final proximal step}\\
\boldsymbol{\beta}_{k}^{r+1} \!\!\! & = & \!\!\! \boldsymbol{\beta}_{k}^{r} + \rho_k(\boldsymbol{\theta}_{k}^{r+1} - \boldsymbol{z}^{r+1}).
\end{eqnarray}
\end{subequations}
The pxADMM shares the same workflow with the cADMM as depicted in Fig.~\subref*{subfig:ADMM}. Criteria for choosing $\rho_k$ and $L_k$ are given in \cite{Hong16}, where the authors also proved under mild conditions that: (1) $\boldsymbol{\theta}_k^{r}$ converge to $\boldsymbol{z}^{r}$ for all $k$; and (2) solution $(\{\boldsymbol{\theta}_{k}^{r}\}, \boldsymbol{z}^{r}, \{\boldsymbol{\beta}_{k}^{r}\})$ converges to a stationary point of Eq.(\ref{P2}).

%
%

The pxADMM reduces the communication overhead in the same way as cADMM does, which was explained in our previous work \cite{Xu19}. However, the proximal step shown in~Eq.\eqref{eq:final proximal step} leads to an inexact, but closed-form solution of the local sub-problem Eq.\eqref{eq:theta minimization} with much cheaper computation cost. Although more iterations may be required towards convergence, the overall computational time can be well reduced. 

\subsection{Privacy Preservation}
\label{subsec:privacy}
Federated learning emphasizes strongly on mobile user’s sole ownership of data and preservation of user privacy. However, recent studies have shown that the shared parameters of the trained models are proved to be vulnerable to disclose sensitive information \cite{Melis18}. Privacy preservation in federated learning can be achieved through various security techniques like secure multi-party computation, homomorphic encryption, and differential privacy. 

To protect the content of each individual piece of trained model, secure multi-party computation involves multiple participants to upload trained models towards the server collaboratively. No matter DNN or GP is used, the distributed gradient descent on user-held training data is protected by secure aggregation with user dropout taken into consideration \cite{bonawitz2016practical}. By exploiting a secure aggregation protocol and a secret-sharing scheme, the privacy of each user-provided model can be guaranteed under an honest-but-curious and active adversarial setting \cite{Bonawitz17}, which supports an arbitrary subset of user dropouts. Other than the above schemes, to verify the correctness of the final aggregation result, a privacy-preserving and verifiable federated learning protocol has been designed with a homomorphic hash function and a secret sharing protocol \cite{Xu20}. However, secure multi-party computation may still leak sensitive information during the learning process.

The key idea of differential privacy in federated learning is to add some “noise” to the trained hyper-parameters with a sensitivity-measured random mechanism, such as Laplace mechanism or Gaussian mechanism \cite{lim2019federated}, which helps mitigate the risk of private information disclosure. However, the injected noise may degrade the performance of the trained model. The feasibility of differential privacy on a client level in federated learning with Gaussian mechanism was demonstrated in \cite{geyer2017differentially}, in which the authors demonstrated the trade-off between the loss of privacy and the modeling performance.

Various homomorphic encryption schemes have been designed to protect the privacy of each mobile user, and their benefits can be summarized as follows: (1) Sensitive information can be held away from the central node; (2) Model accuracy can be kept intact \cite{Phong18}. With the additive homomorphic encryption, stochastic gradient descent can be protected against an honest-but-curious central node. Another hybrid scheme combining the differential privacy technique and the threshold homomorphic encryption was designed, which can further resist collusion attacks between the colluding server and participants \cite{Truex19}. 

Note that in our work, no matter which distributed model training method is involved, homomorphic encryption is preferred to be used by the mobile users to upload hyper-parameter estimates or local gradients towards the central node. Among others, a representative homomorphic encryption method that suits our desired applications can be identified in \cite{Pascal99} with more implementation details disclosed in \cite{Yang19}. 

\subsection{FedLoc: A New Umbrella of Old Modules}
In the previous sections, we have introduced two important classes of learning models, namely the deep neural network models and Gaussian process models, and a few distributed hyper-parameter optimization schemes tailored to these two models, as well as the state-of-the-art privacy preservation methods for mobile data. These constitute the major ingredients of a novel cooperative, data-driven, learning model-based framework for localization and location data processing.  

For clarity, we give a complete procedure of the FedLoc framework in Algorithm~1, which can be adopted for both the cooperative localization and the cooperative location data processing. Various live use cases in different application sectors already fall into or can be revised to suit our FedLoc framework. In Section~\ref{sec:use-cases}, we will show a few representative use cases and survey some related works that can be made adapt to the FedLoc. 

\begin{algorithm}[t!]
\caption{FedLoc Framework under Cloud-Based Network Infrastructure}
	\KwIn{(1) A massive number of collaborating mobile terminals with index $k=1,2,...,K$; (2) Local data $\mathcal{D}_k = \{\boldsymbol{X}_k, \boldsymbol{y}_k\}$, where the inputs and outputs are positions/position related measures; (3) A learning model, for instance a DNN or a GP model.}
	\KwOut{Optimal hyper-parameters $\boldsymbol{\theta}^{*}$ of the global learning model.}
	
	\textbf{Initialization}: Initial hyper-parameters of the selected learning model, $\boldsymbol{\theta}^0$; iteration index, $\eta = 0$. \\

	\For{(outer iterations) $\eta=0,1,...$}{
	    1. The core network sends a probing signal to all mobile terminals and identifies which ones are idle during this round. The idle terminals form a set, $\mathcal{K}_{\eta}$. 
	    
	    2. The core network sends a seed to the selected terminals for encoding the messages as well as the current hyper-parameter estimate, $\boldsymbol{\theta}^{\eta}$. 
	    
		\For{(inner iterations) each idle mobile terminal $k \in \mathcal{K}_{\eta}$ in parallel}{
			
			1. Use the local data, $\mathcal{D}_k$, or a fraction of it to update the hyper-parameter $\boldsymbol{\theta}_{k}^{\eta+1}$, for instance, via FedAvg/FedProx for DNN or via cADMM/proximal ADMM for GP.
			
			2. Encrypt the local update of the global model hyper-parameters as a message using for instance homomorphic encryption.
			  
            3. Send the encrypted message to the core network. 
			
		}
		
		3. The core network receives all encrypted messages from the mobile terminals indexed in $\mathcal{K}_{\eta}$ and performs decryption.
		
		4. The core network updates the global learning model hyper-parameters via consensus.
				
		5. Finish this round and reset $\eta = \eta+1$.
		
		6. Repeat the above iterations (1)-(5) until certain stopping criteria are satisfied. 
	}
	
	The approximated global hyper-parameters is $\boldsymbol{\theta}^{*} = \boldsymbol{\theta}^{\eta}$.
\end{algorithm}
\begin{table*}[t!]
\centering
\begin{tabular}{| p{3.5cm} | p{2.6cm} | p{2.6cm} | p{3cm} | p{3cm} |}
\hline
\textbf{Wireless infrastructure} & \textbf{Max uplink data rate (Mbps)} & \textbf{Max downlink data rate (Mbps)} & \textbf{Number of DNN weights (Million)} &\textbf{Configuration} \\
\hline 
5G \cite{5Gpeak} &10,000 & 20,000 & 312.5 & IMT-2020 peak rate \\
\hline
4G \cite{4G} & 500 & 1000 & 15.625 & IMT-advanced \\
\hline
WiFi-6(ax) \cite{Wifi6} & 2400 & 2400 & 75 & 160MHz 2*2MIMO 1024-QAM 802.11ax \\
\hline
WiFi-5(ac) \cite{Wifi5} & 1733 & 1733 & 54.16 &  160MHz 2*2MIMO 256-QAM 802.11ac \\
\hline
\end{tabular}
\vspace{0.2cm}
\caption{Downlink and uplink data rate of different wireless infrastructures under specific configurations and the number of hyper-parameters of a selected learning model (taking the DNN weights as example) that can be supported. The number of the DNN weights (in million) shown in the fourth column is equal to the uplink rate (given in the second column) divided by 32 bits per DNN weight.}
\label{tab:vol-wireless-infrastructure}
\end{table*}

\section{Network Infrastructures for FedLoc}
\label{sec:NetInf}
As it is widely known, federated learning needs to communicate a big number of model parameters continuously over the air, especially when DNN is adopted as the learning model. In this section, we introduce two promising network infrastructures to meet the communication requirements of the proposed FedLoc framework. Specifically, a cloud-based wireless network infrastructure is introduced in Subsection~\ref{subsec:cloud-based}, while an emerging edge-based one is introduced in Subsection~\ref{subsec:edge-based}. More fresh discussions on using parallel infrastructures to support scalable learning paradigms for data-driven wireless applications can be found in our recent work \cite{xu2020scalable}.
\begin{figure}[t]
	\includegraphics[width=0.48\textwidth]{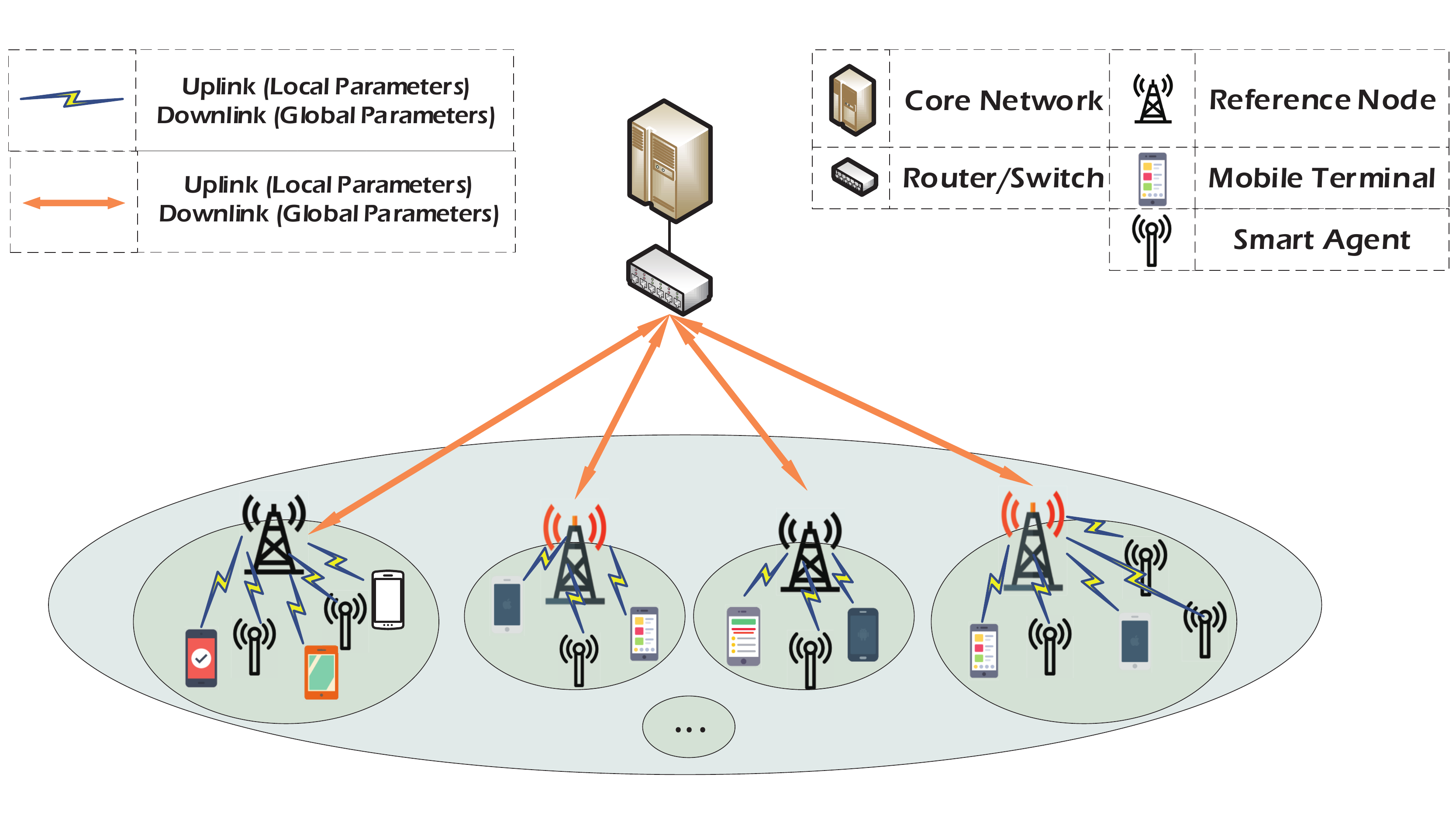}
	\caption{Cloud-based network infrastructure for supporting the proposed FedLoc framework. For illustration purpose only, the whole deployment area is divided into many non-overlapping sub-areas, and for each sub-area there is a bunch of mobile terminals willing to collaborate.}
	 \label{fig:figure1a}
\end{figure}

\subsection{Cloud-based Infrastructure}
\label{subsec:cloud-based}
For ease of understanding, a complete picture of the network infrastructure is depicted in Fig.~\ref{fig:figure1a} for learning model-based cooperative localization. The key elements of this network as well as their functionality are summarized as follows:
\begin{enumerate}
\item \textit{Reference Network Node} is equipped with cache, storage, and communication entities. A reference network node communicates with the mobile terminals deployed in its communication range to exchange learning model related information. Both the position and the transmit power of a reference network node are assumed to be precisely known. Representative reference network nodes include 5G macro and micro base stations, WiFi access points, BLE beacons, etc. Especially the emerging 5G and WiFi-6 network are able to provide low-latency, high throughput wireless transmission to FedLoc, which requires to transmit a big amount of model parameters in every iteration. Table~\ref{tab:vol-wireless-infrastructure} gives some numbers. For better intuitions, two specific examples are given below. The 5G network with the highest throughput can support a $9$-layer fully-connected DNN with the network layout ``20000-30000-10000-100000-10000-10000-10000-1000-10'' that has around $300$ million weights. The 4G network, however, can only support an $8$-layer fully-connected DNN with a much smaller network layout ``5000-5000-10000-3000-9000-2000-200-10'' with around $15$ million weights.
\item \textit{Mobile Terminal (MT)} is equipped with sensing, logging, computing, storage, and communication entities. Moreover, the MT has installed the designated mobile applications for carrying out the calibration work. The MT collects position related measurements, obtains a local update of the global learning model parameters, and uploads them to the core network. All the computations are conducted on-device using the local data only. Here, the mobile terminal refers to a smartphone specifically. It is noteworthy that modern smartphones are equipped with a basket of inertial sensors, including accelerometer, gyroscope, magnetometer, barometer, pedometer, barcode/QR code sensors, that can be exploited for localization or localization-related tasks. Apart from the rapid development of the hardware, a number of mobile machine learning platforms are under development, such as Tensorflow by Google, Core ML by Apple, Caffe2 by Facebook, Paddle Lite by Baidu, MNN by Alibaba, etc. Mobile users can easily deploy different deep learning models on their smartphones in the near future.
\item \textit{Fixed Smart Agents} are equipped with sensing, logging, computing, storage and communication entities. Representative smart agents include IoT machines, wireless sensors, robots, smart traffic lights, unmanned aerial vehicles (UAVs), micro-base stations that are collecting location data continuously. 
%
\item \textit{Core Network} is equipped with high-speed computing, cache/storage and communication entity. The local updates from the mobile users are aggregated to the core network to compute a global parameter update. After the training phase is over, the approximated global learning model will be stored in the core network and used for predicting a new position in the online phase. Since the heavy computations have been offloaded to a number of mobile users, the core network can perform smarter coordination of different tasks and resources, so as to make the whole network agile and adaptive to the fast changing environments. 
\end{enumerate}

\subsection{Edge-based Infrastructure}
\label{subsec:edge-based}
\begin{figure}[t]
	\includegraphics[width=0.48\textwidth]{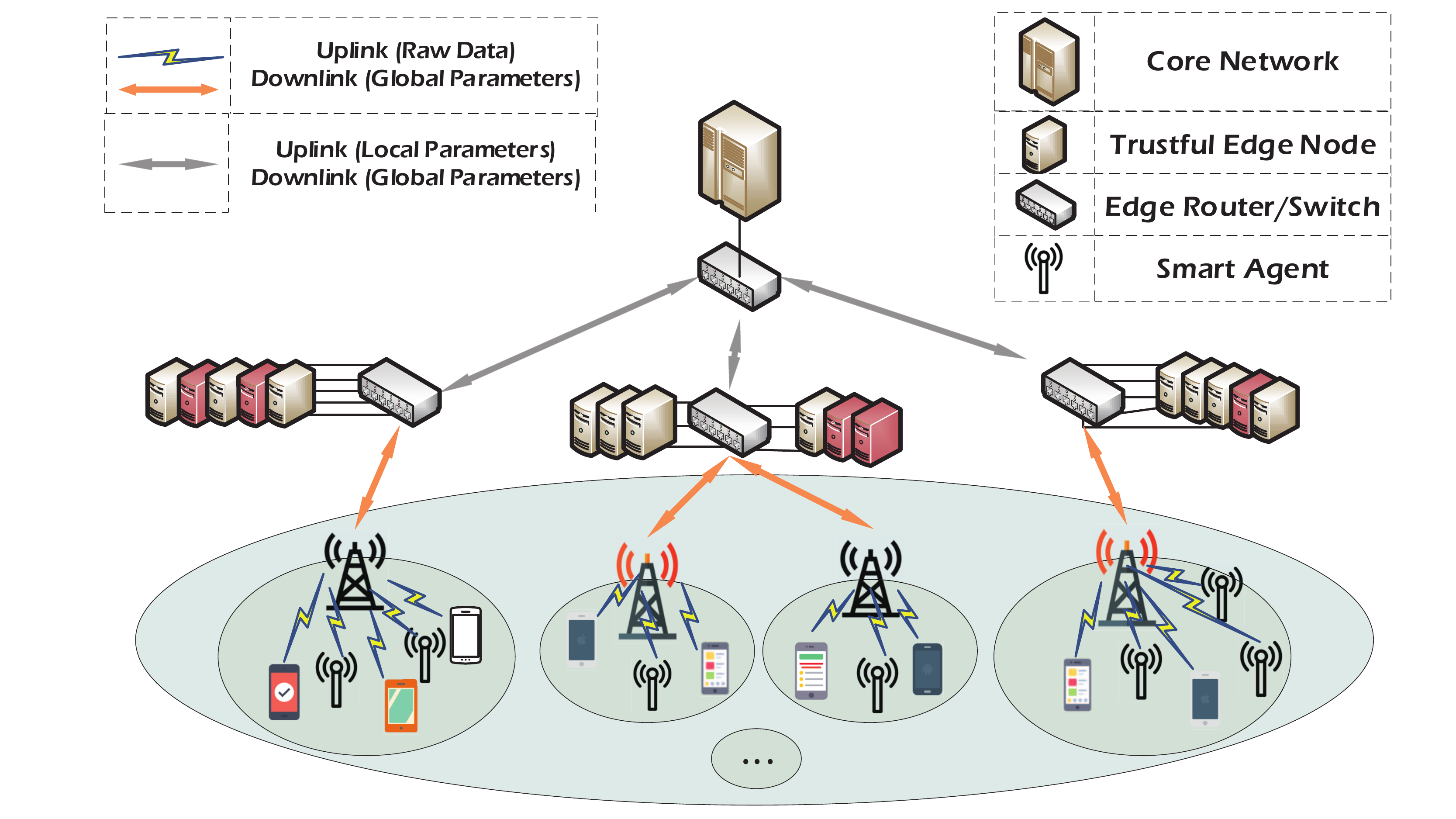}
	\caption{Edge-based network infrastructure for supporting the FedLoc framework.}
	 \label{fig:figure1b}
\end{figure}

In the second infrastructure,  the mobile users or smart agents can upload their local data to a trustful third-party edge node, where there is sufficient storage and computation power for handling learning tasks. For clarity, we show this network infrastructure in Fig.\ref{fig:figure1b}. The edge node first pre-processes the received data and then offloads the model fitting task to a number of computing units. Each edge node is in charge of building a locally-global learning model and transmits the trained hyper-parameters to the core network for consensus and coordinated control. This infrastructure is more suitable for building a number of regional global models for location data processing. The third use case that we will show in the next section can potentially benefit a lot from this edge-based infrastructure. 

\section{Use Cases of FedLoc}
\label{sec:use-cases}

This section aims to shed more light on the FedLoc framework with various live use cases. In particular, we showcase: (1) DNN-based static localization/fingerprinting; (2) DNN-based smartphone sensor calibration for accurate navigation with low-sampling-rate GPS; (3) GP-based state-space model for target tracking and navigation; and (4) GP-based wireless traffic prediction in 5G C-RAN. The first three use cases relate to localization, while the last one relates to location data processing and prediction. Most of the above uses cases are summarized from our recent works. We also survey related works that can easily fit into the FedLoc framework.
%
%

\subsection{DNN-Based Static Localization/Fingerprinting}
There exist various statistical methods using wireless measurements, such as ToA, RSS, proximity \cite{GG09, Yin15}, for static target localization. These methods rely on empirical propagation models. In this subsection, we show a different static localization method using DNN, which can benefit from the federated learning framework. DNN-based localization is preferred for complex indoor wireless environments, for which sophisticated empirical models are either not available or incapable of capturing the underlying propagation mechanism. 

\begin{figure}[t]
	\includegraphics[width=0.46\textwidth]{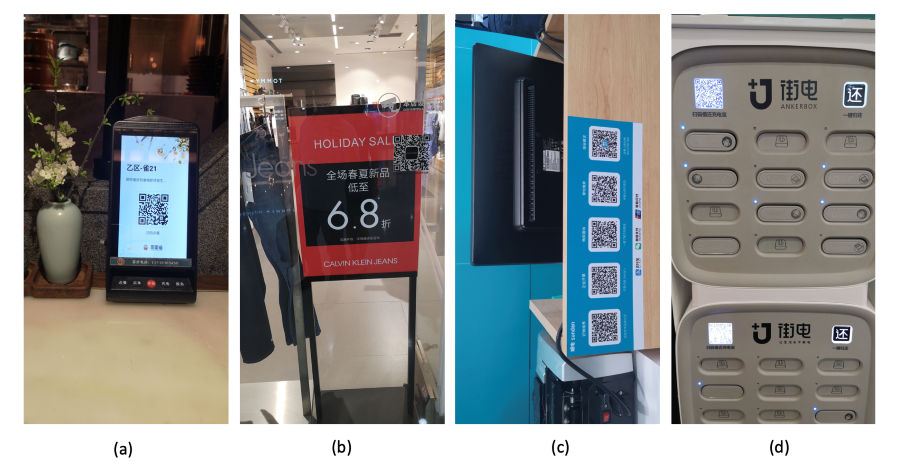}
	\caption{All the QR labels were photoed in a modern shopping mall in Shenzhen, China. (a) QR codes for ordering foods for a specific dining table; (b) QR code for promotion information at a shop; (c) QR codes for various different services, including product recommendation, payment, etc at the cashier of a shop; (d) QR code for renting mobile power bank.}
	 \label{fig:figure2}
\end{figure}

Let us take a look at three representative indoor scenarios:
\begin{itemize}
\item \textbf{Indoor shopping mall}, where there are a bunch of WiFi/BLE access points and micro base stations for public data traffic. In addition, thanks to the rapid spread of 5G for IoT and machine-type communications (MTC), there are now a large number of machines/landmarks with QR codes in the shops. By scanning the QR codes, customers can easily get shopping mall information and promotional information. Some live examples are demonstrated in Fig.~\ref{fig:figure2}.
\item \textbf{Indoor museum}, where there are a bunch of WiFi/BLE access points in the exhibition rooms and a considerable number of QR labels attached to the exhibits to serve as references. Similarly, by scanning the QR codes a visitor can easily get access to detailed interpretation of the exhibits on his/her mobile terminal. 
\item \textbf{Indoor office}, where there are a bunch of WiFi/BLE access points in the whole office area, and a large number of QR labels are placed on all valuable assets in the room. 
\end{itemize}
\begin{figure}[t]
	\includegraphics[width=0.5\textwidth]{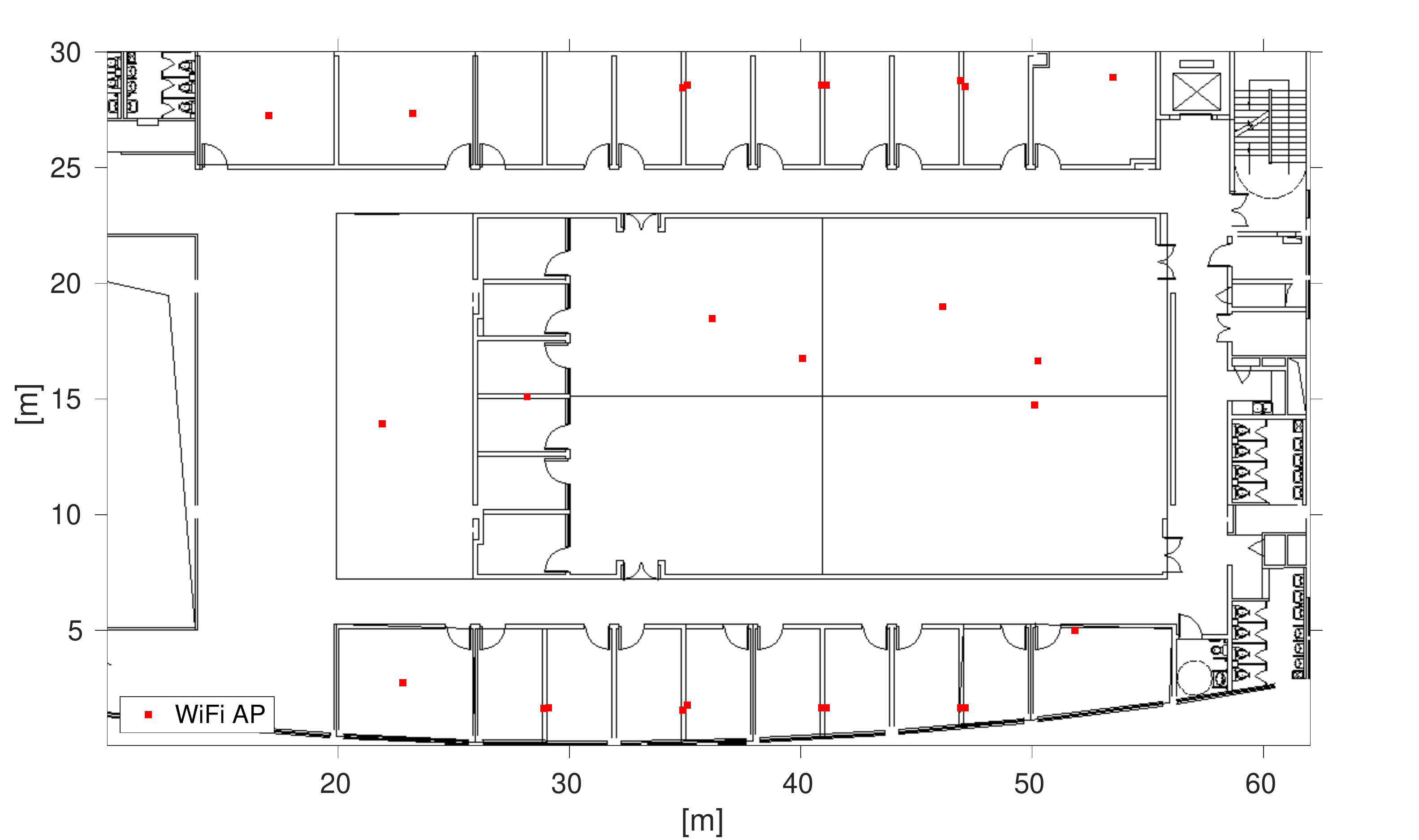}
	\caption{A typical indoor office environment at the Chinese University of Hong Kong (Shenzhen), where two dozens of WiFi access points are deployed in the offices and laboratories. For this conceptual example, an input, $\boldsymbol{x}_i$, is a vector of $P=26$ RSS values, and the corresponding output, $\boldsymbol{y}_i = [p^{x}_{i}, p^{y}_{i}]$ is a 2D position.}
	 \label{fig:figure3}
\end{figure}

The DNN-based static localization/fingerprinting needs to be trained with a big data set $\mathcal{D}$, where the training input, $\boldsymbol{X}$, contains the radio features at different locations and the training output, $\boldsymbol{y}$ contains the corresponding locations. As a concrete example, we assume that a training input comprises RSS measured with respect to $P$ WiFi/BLE access points, $\boldsymbol{x}_i = [RSS_{i,1}, RSS_{i,2},...,RSS_{i,P}]$, and the output $\boldsymbol{y}$ is a position (2D or 3D) at which the radio feature is measured. More sophisticated measurements such as magnetic fields and channel state information (CSI) can be used instead of the RSS or jointly used with the RSS. Note that an output $\boldsymbol{y}_i$ is either measured precisely at the calibration points by paid workers or imprecisely (for instance, with the aid of the landmark points and manual click on the indoor map displayed on the mobile application) by voluntary users. In either case, we assume the output is subject to additional independent noise. A concrete example is illustrated in Fig.~\ref{fig:figure3}. 

\begin{figure*}[t]
	\includegraphics[width=0.98\textwidth]{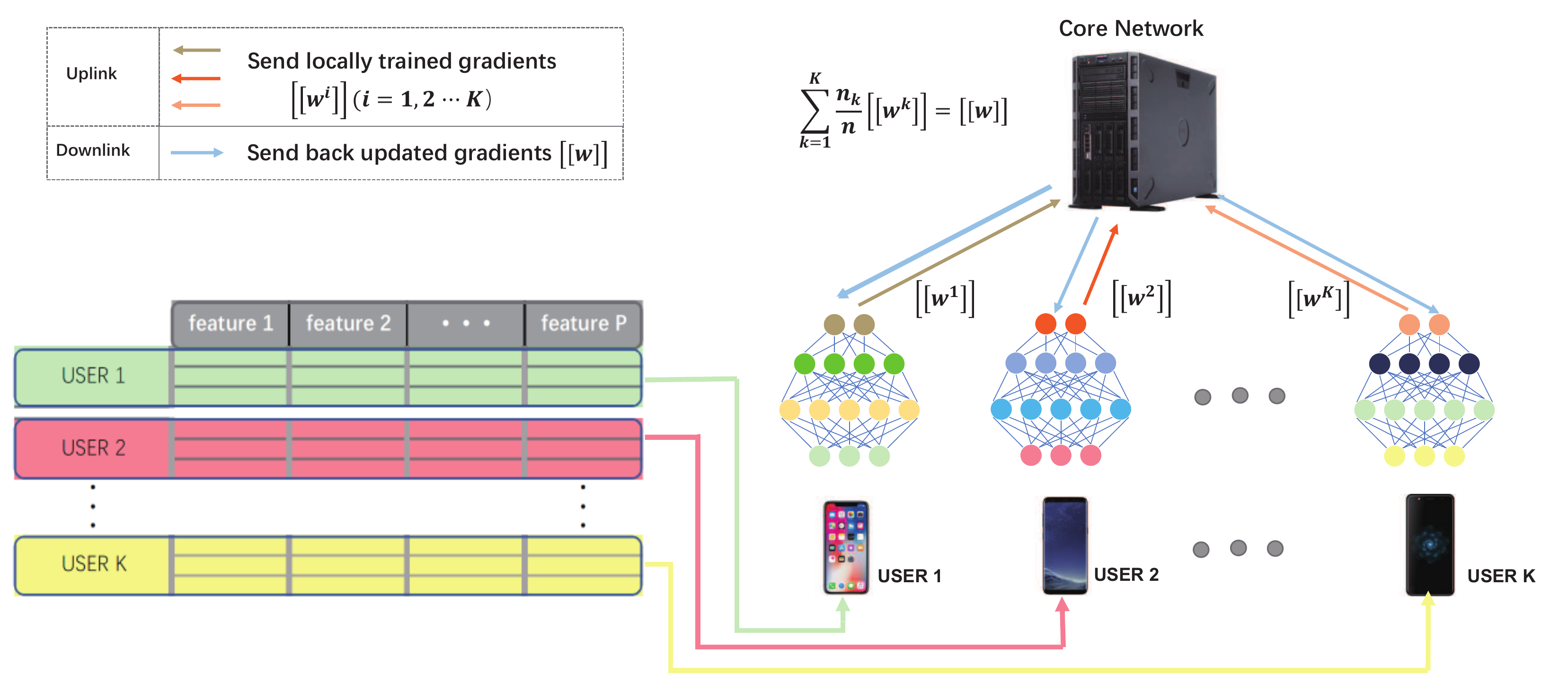}
	\caption{Illustration of DNN-based static localization. Here, $[[ W ]]$ represents encrypted NN weight parameters using for instance Homomorphic Encryption (HE). The $P$ features are RSS values collected from the WiFi access points in the deployed area.}
	 \label{fig:figure5}
\end{figure*}

The regression problem can be formulated as
\begin{equation}
\boldsymbol{y}_i = f(\boldsymbol{x}_i; \boldsymbol{\theta}) + \boldsymbol{n}_i,
\end{equation}
where $f(\boldsymbol{x}; \boldsymbol{\theta}): \mathbb{R}^{dx} \rightarrow \mathbb{R}^{dy}$ represents a DNN with an input of $dx=P$ features and the neural network weights $\boldsymbol{\theta}$ to be tuned. The regression function $f(\boldsymbol{x}; \boldsymbol{\theta})$ is also known as RSS map or fingerprinting map in the literature.

In order to adopt the federated learning framework, we deploy a large number of mobile terminals, and each is responsible for a particular area, possibly overlapping with its neighboring areas. The $k$-th mobile terminal collects a data set $\mathcal{D}_k = \{ \boldsymbol{X}_k, \boldsymbol{y}_k \}$ and uses it to train a local update of the global parameters. Concretely, each mobile user solves
\begin{equation}
\boldsymbol{\theta}_k = \arg \min_{\boldsymbol{\theta}} \sum_{ \forall \{ \boldsymbol{x}_i, \boldsymbol{y}_i \} \in \mathcal{D}_k} \!\!\!\!\!\! || \boldsymbol{y}_i - f(\boldsymbol{x}_i; \boldsymbol{\theta}) ||_{2}^2.
\end{equation}
All the mobile terminals cooperate to perform Algorithm 1. Since in this use case, the global objective is readily in the form of summation, therefore we can set the weights $\beta_k$ to be the ratio $\vert \mathcal{D}_k \vert /\sum_{j \in \mathcal{K}_\eta} |D_j| $ in the $\eta$-th iteration and update $\boldsymbol{\theta}^{\eta+1} = \sum_{k \in \mathcal{K}_{\eta}} \beta_k \boldsymbol{\theta}_{k}^{\eta}$. When the messages are exchanged between the core network and mobile terminals, they are first encrypted by the mobile terminals and decrypted in the core network using homomorphic techniques. The workflow of the FedLoc for DNN-based static localization is shown in Fig.~\ref{fig:figure5}. 

After the training procedure is terminated, the central node will obtain an approximated global estimate of the hyper-parameters, denoted by $\hat{\boldsymbol{\theta}}$. Given a new vector of RSS measurements, $\boldsymbol{x}_* = [RSS_{*,1}, RSS_{*,2},...,RSS_{*,P}]$, reported to the central node, the trained learning model will map it then to the desired position estimate through $\boldsymbol{p}_{*} = f(\boldsymbol{x}_{*}; \hat{\boldsymbol{\theta}})$. 

Various works on using deep learning models and RSS measurements for indoor fingerprinting have been published in recent years, for instance \cite{Han18, Liu18RSS, Hsieh19, Wang20} based on DNN, CNN, LSTM. Although these works are originally centralized algorithms, they can be implemented in a distributed manner under our FedLoc framework.


\subsection{DNN-Based Vehicle Navigation with Low Sampling Rate GPS}
For land vehicle navigation, combining the inertial measurement unit (IMU) and global positioning system (GPS) embedded in a smartphone is still the main-stream technical solution. The GPS can readily provide accurate vehicle positions when the majority of the satellite signals are in line-of-sight (LOS) propagation with relatively high RSS. On the other hand, the IMU assembles, primarily, a three-axis acceleration sensor and a three-axis gyroscope, to determine the position and velocity of a vehicle. The main functionality of the IMU is to provide vehicle positions with a much higher sampling rate ($> 50$ Hz) between two consequent GPS position estimates (with 1 Hz by default). Unfortunately, when a vehicle enters into certain areas with severe signal blockage, the received GPS signal will be very weak or even undetectable, leading to significantly degraded position estimate. On the other hand, solely relying on low-end IMU measurements for high-accuracy navigation is impractical due to the sensor bias, scale-factor error, and other random errors that accumulate over time. How can we maintain a satisfactory positioning accuracy for the case that GPS signals are occasionally available for harsh wireless environments, such as in the city center or forest? We demand a smart solution with affordable computational complexity. 

Towards this end, we introduce in this subsection a machine learning-based approach that can be implemented on commercial smartphones and is able to provide high navigation accuracy using low-end inertial sensors and low-sampling-rate GPS. Inertial sensors are used to continuously estimate the vehicle velocity and position at higher sampling rate, while low-sampling-rate GPS signals are used for IMU calibration occasionally (for example every 60 seconds). When the GPS signal is not available, we use pre-trained DNNs to calibrate the inertial sensor errors.

To be concrete, we adopt two DNNs to estimate/predict the velocity $v_{t, NN1}$ and the yaw angle $y_{t, NN2}$ of the vehicle, respectively. In the model training phase, both DNNs take measurements from the smartphone inertial sensors as the input while the GPS velocity and yaw angle measurements are taken as the outputs/labels.

The first DNN takes the following inputs:
\begin{itemize}
\item The velocity $\widetilde v^n_t = ((v_t^{nx})^2+(v_t^{ny})^2+(v_t^{nz})^2)^{1/2}$ calculated from the inertial sensor data;
\item The sequence of angular velocity $\{\omega ^{bz}_{t-l} ,..., \omega ^{bz}_{t}\}$ of the vehicle;
\item The sequence of smoothed linear acceleration along the front direction of the vehicle, denoted as $\{a^{nx}_{t-l} , ..., a^{nx}_{t}\}$.
\end{itemize}
The DNN output is the velocity $v_{t,NN1}$ set to be the GPS velocity $v_{t, GPS}$ as the ground-truth in the training data set. 

Similarly, the second DNN takes the following inputs:
\begin{itemize}
\item The sequence of smoothed linear acceleration, denoted as $\{a_{by} ,...,a_{by}\}$;
\item The sequence of angular velocity $\{\omega ^{bz}_{t-l} ,..., \omega ^{bz}_{t}\}$;
\item The compensated yaw sequence $\{y_{t-l}, ..., y_t \}$.
\end{itemize}
The DNN output is the yaw angle $y_{t, NN2}$ set to the GPS yaw angle $y_{tGPS}$ as the ground-truth in the training data set.

Our recent work in \cite{Qi19} presented a centralized implementation, where interested readers can find more details about the measurements, configurations of the DNNs, as well as a diagram of the whole navigation system. In this paper, we are interested in designing a distributed counterpart. To this end, we let the two DNNs be trained individually by a batch of collaborating mobile users according to Algorithm~1 with the DNN weights optimized using either the FedAvg algorithm or the FedProx algorithm. The information exchange procedure remains the same as the first use case. In the online use phase, the two DNNs will calibrate the inertial sensor error aggregation when there is no GPS signal at hand. Some primary results for this use case will be shown in Section~\ref{sec:results}.

\subsection{GP-Based State-Space Model (GPSSM) for Target Tracking}
State-space models (SSM) are outstanding for modeling a time series $\boldsymbol{y}_{1:T}\triangleq\{\boldsymbol{y}_t\}_{t=1}^T$ with latent states $\boldsymbol{x}_{0:T}\triangleq\{\boldsymbol{x}_t\}_{t=0}^T$. An SSM comprises a transition function, $f(\boldsymbol{x}):\mathbb{R}^{dx}\rightarrow\mathbb{R}^{dx}$ and a measurement function, $g(\boldsymbol{x}): \mathbb{R}^{dx}\rightarrow\mathbb{R}^{dy}$. Concretely, an SSM is given by
\begin{align}
\boldsymbol{x}_t & = f(\boldsymbol{x}_{t-1}) + \boldsymbol{e}_{t-1}, \nonumber \\
\boldsymbol{y}_t & = g(\boldsymbol{x}_t) + \boldsymbol{n}_t, 
\label{eq:ssm}
\end{align}
where $\boldsymbol{x}_t \in \mathbb{R}^{dx}$ is the latent state, $\boldsymbol{y}_t \in \mathbb{R}^{dy}$ is the measurement, $\boldsymbol{e}_{t}$ is the process noise, and $\boldsymbol{n}_t$ is the measurement noise at time instance $t$, respectively. Traditional SSM restricts both the transition function $f(\boldsymbol{x})$ and the measurement function $g(\boldsymbol{x})$ to empirical, parametric functions \cite{Bar-Shalom2001}, whose parameters can be learned through the expectation-maximization (EM) algorithm \cite{Schon11} or Markov chain Monte Carlo (MCMC) algorithm \cite{Andrieu10}. 

Since GP models provide outstanding performance in function approximation with a natural and inherent uncertainty region, they have been adopted to model complicated nonlinear functions in the SSMs, leading to the GPSSM \cite{Frigola14a}. Early variants of the GPSSM were learned by finding the maximum \textit{a posteriori} (MAP) estimates of the latent states, generating various successful positioning applications, among others the RSS-based WiFi localization \cite{Ferris07}, the human motion capture \cite{Wang08}, and the IMU-based slotcar tracking \cite{Ko11}, etc. The first fully probabilistic learning procedure of the GPSSM was proposed in \cite{Frigola13} using particle Markov Chain Monte Carlo (PMCMC). In order to reduce the heavy computational load of the sampling method used in \cite{Frigola13}, a number of different variational learning procedures were developed in \cite{Frigola14a, Eleftheriadis17, Ialongo18a, Ialongo18b} upon the classical variational sparse GP framework \cite{Titsias09}. 

\begin{figure}[t]
	\centering
	\footnotesize
	\begin{tikzpicture}[align = center, latent/.style={circle, draw, text width = 0.5cm}, observed/.style={circle, draw, fill=gray!20, text width = 0.5cm}, transparent/.style={circle, text width = 0.5cm}, node distance=1.1cm]
	\node[latent](x0) {$\boldsymbol{x}_0$};
	\node[latent, right of=x0](x1) {$\boldsymbol{x}_{1}$};
	\node[transparent, right of=x1](x2) {$\cdots$};
	\node[latent, right of=x2](xt-1) {$\!\!\boldsymbol{x}_{t-1}\!\!$};
	\node[latent, right of=xt-1](xt) {$\boldsymbol{x}_{t}$};
	\node[transparent, right of=xt](xinf) {$\cdots$};
	\node[transparent, above of=x0](f0) {$\cdots$};
	\node[latent, above of=x1](f1) {$\boldsymbol{f}_{1}$};
	\node[transparent, right of=f1](f2) {$\cdots$};
	\node[latent, above of=xt-1](ft-1) {$\!\!\boldsymbol{f}_{t-1}\!\!$};
	\node[latent, above of=xt](ft) {$\boldsymbol{f}_{t}$};
	\node[transparent, right of=ft](finf) {$\cdots$};
	\node[transparent, below of=x0](g0) {$\cdots$};
	\node[latent, below of=x1](g1) {$\boldsymbol{g}_{1}$};
	\node[transparent, below of=x2](g2) {$\cdots$};
	\node[latent, below of=xt-1](gt-1) {$\!\!\boldsymbol{g}_{t-1}\!\!$};
	\node[latent, right of=gt-1](gt) {$\boldsymbol{g}_{t}$};
	\node[transparent, right of=gt](ginf) {$\cdots$};
	\node[observed, below of=g1](y1) {$\boldsymbol{y}_{1}$};
	\node[transparent, below of=g2](y2) {$\cdots$};
	\node[observed, below of=gt-1](yt-1) {$\!\!\boldsymbol{y}_{t-1}\!\!$};
	\node[observed, right of=yt-1](yt) {$\boldsymbol{y}_{t}$};
	\node[transparent, right of=yt](yinf) {$\cdots$};
	\draw[-latex] (x0) -- (f1);
	\draw[-latex] (f1) -- (x1);
	\draw[-latex] (x1) -- (f2);
	\draw[-latex] (ft-1) -- (xt-1);
	\draw[-latex] (xt-1) -- (ft);
	\draw[-latex] (ft) -- (xt);
	\draw[-latex] (xt) -- (finf);
	\draw[-latex] (x1) -- (g1);
	\draw[-latex] (xt-1) -- (gt-1);
	\draw[-latex] (xt) -- (gt);
	\draw[-latex] (g1) -- (y1);
	\draw[-latex] (gt-1) -- (yt-1);
	\draw[-latex] (gt) -- (yt);
	\draw[ultra thick]
		(f0) -- (f1)
		(f1) -- (f2)
		(f2) -- (ft-1)
		(ft-1) -- (ft)
		(ft) -- (finf)
		(g0) -- (g1)
		(g1) -- (g2)
		(g2) -- (gt-1)
		(gt-1) -- (gt)
		(gt) -- (ginf);
	\end{tikzpicture}
	\caption{Graphical representation of GPSSM. The shaded nodes represent the measurements, while the transparent nodes represent the latent variables. Variables belonging to the same GP are connected by a thick edge.}
	\label{fig:figure6}
\end{figure}
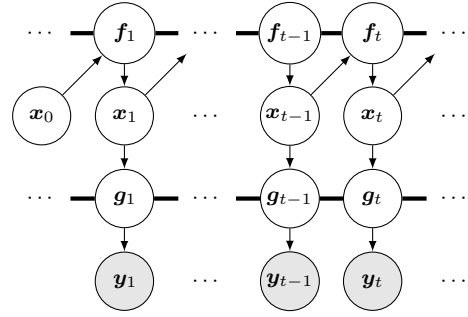

A general GPSSM can be formulated as 
\begin{align}
f(\boldsymbol{x}) & \sim  \mathcal{GP}(m_f(\boldsymbol{x}), k_f(\boldsymbol{x}, \boldsymbol{x}'; \boldsymbol{\theta}_f)), \nonumber \\
g(\boldsymbol{x}) & \sim  \mathcal{GP}(m_g(\boldsymbol{x}), k_g(\boldsymbol{x}, \boldsymbol{x}'; \boldsymbol{\theta}_g)), \nonumber \\
\boldsymbol{x}_0 & \sim p(\boldsymbol{x}_0), \nonumber \\
\boldsymbol{f}_t & = f(\boldsymbol{x}_{t-1}), \nonumber \\
\boldsymbol{x}_t | \boldsymbol{f}_t & \sim \mathcal{N}(\boldsymbol{x}_t | \boldsymbol{f}_t, \boldsymbol{Q}), \nonumber \\
\boldsymbol{g}_t & = g(\boldsymbol{x}_t), \nonumber \\
\boldsymbol{y}_t | \boldsymbol{g}_t & \sim \mathcal{N}(\boldsymbol{y}_t | \boldsymbol{g}_t, \boldsymbol{R}), 
\end{align}
with the model hyper-parameters $\{\boldsymbol{\theta}_f, \boldsymbol{\theta}_g, \boldsymbol{Q}, \boldsymbol{R}\}$, where $\boldsymbol{\theta}_f$ and $\boldsymbol{\theta}_g$ are the kernel hyper-parameters of the GPs, $\boldsymbol{Q}$ and $\boldsymbol{R}$ are the covariance matrices of the process noise and the measurement noise, respectively. For clarity, Fig.~\ref{fig:figure6} shows a graphical representation of the GPSSM. In the following, we will first introduce the standard GPSSM, which requires a big set of calibrated data to train both the transition function $f$ and the measurement function $g$. Then, we will briefly mention the advanced variational GPSSM proposed initially in \cite{Frigola14a}.  

We start with the transition function of the standard GPSSM. The GP regression model for the transition function, $f$, is $\boldsymbol{x}_{t+1} = f(\boldsymbol{x}_t) + \boldsymbol{e}_t$, where the output $\boldsymbol{x}_{t+1} \in \mathbb{R}^{dx}$ is the state at time $t+1$, $\boldsymbol{x}_{t} \in \mathbb{R}^{dx}$ is the current state at time $t$, the unknown function $f(\boldsymbol{x}_t) : \mathbb{R}^{dx} \rightarrow \mathbb{R}^{dx}$ is essentially a multi-output GP \cite{RW06}, and $\boldsymbol{e}_t$ is a vector of noise terms. For simpler implementation, we could model each entry of the state, say the $j$-th, by an independent GP as $[\boldsymbol{x}_{t+1}]_{j} = f_{j}(\boldsymbol{x}_t) + e$, where $f_{j}(\boldsymbol{x}_t) : \mathbb{R}^{dx} \rightarrow \mathbb{R}$ is now a single-output GP. As discussed in Section~\ref{sec:lmodels}, we need to select a kernel function, $k_{f}(\boldsymbol{x}_t, \boldsymbol{x}_{t'}; \boldsymbol{\theta})$ to represent the correlation between the states at different time instances. When the input dimension is small/modest, using the ARD kernel is often a good choice. While for large input dimension, advanced kernels such as the arc-cosine kernel and the NTK should better be tried out. 

The above GP models can be trained with a data set of calibrated trajectories, $\mathcal{D}_j \triangleq \{\boldsymbol{X}, \tilde{\boldsymbol{x}}_j \}$, where $\tilde{\boldsymbol{x}}_j  = [[\boldsymbol{x}_{1}]_{j}, [\boldsymbol{x}_{2}]_{j}, ..., [\boldsymbol{x}_{T}]_{j}]^T$ is a vector of outputs and $\boldsymbol{X}=[\boldsymbol{x}_0, \boldsymbol{x}_1,...,\boldsymbol{x}_{T-1}]^T$ is a matrix of inputs. One could follow Eq.(\ref{eq:GP-global-obj}) to solve for the global ML hyper-parameter estimate. To implement the FedLoc framework, one could let $K$ mobile users collaborate to approximate the global ML hyper-parameter estimate according to Eq.(\ref{eq:GP-local-approx}) with the local trajectories walked by each individual. The central node makes consensus on the local hyper-parameter estimates.

The GP regression model for the measurement function is $\boldsymbol{y}_{t} = g(\boldsymbol{x}_t) + \boldsymbol{n}_t$, where the input $\boldsymbol{x}_{t} \in \mathbb{R}^{dx}$ is the state at time $t$, the output $\boldsymbol{y}_t$ is a vector of wireless measurements, and the unknown function $g(\boldsymbol{x}_t) : \mathbb{R}^{dx} \rightarrow \mathbb{R}^{dy}$ is essentially another multi-output GP. Similar to the modeling of the transition function, we apply an independent GP for each single entry of the output. Training the measurement function is similar to that of the transition function, $f$, introduced above. Interested readers can find more details about using GPs to model $f$ and $g$ in \cite{Zhao16, Yin17}. After the GPSSM is built, it can be combined with the celebrated particle filter or smoother \cite{Gustafsson2010} to reconstruct unknown trajectories. In \cite{Xie19}, we proposed a practical real indoor navigation system prototype based on the GPSSM and achieved improved navigation accuracy in various tests with smartphone sensory data. Moreover, we derived both the posterior- and parametric Cramer-Rao bounds for general nonlinear filtering problems based on GPSSM in \cite{Zhao19}.  

One drawback of the above standard GPSSM lies in the need for a relatively large training data set with calibrated latent states, which requires a large amount of labor force. To remedy this drawback, some recent works \cite{Frigola14a, Ialongo18a} incorporated the variational inference technique \cite{Titsias09} into the standard GPSSM to jointly estimate the GPSSM model hyper-parameters and the latent states on the fly. The variational GPSSM does not require a historical calibrated data set, but as tradeoff it has to deal with a large-scale optimization problem. In order to make it adapt to the FedLoc framework, one may consider using the distributed variational inference techniques \cite{Gal14} with the GPSSM. 

\begin{figure}[t]
\centering
\includegraphics[trim=20 20 30 20, clip, width=7cm]{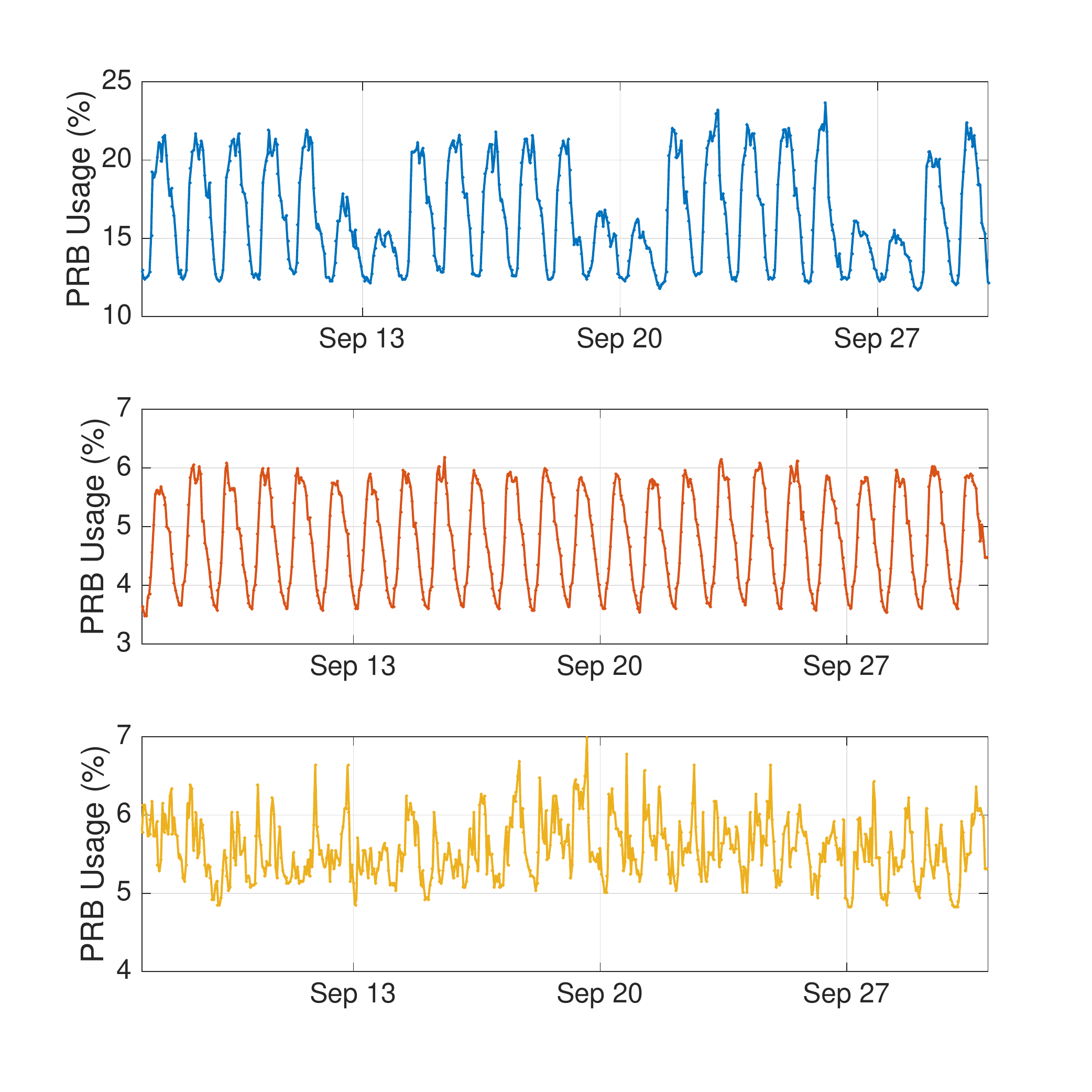}
\caption{The PRB usage curves of three base stations collected in three southern cities of China in 30 days. The data profile in the first panel reflects a typical office area, in which the traffic pattern shows a strong weekly periodic trend in accordance with weekdays and weekends. The data profile in the second panel reflects a typical residential area, in which the traffic pattern shows a strong daily trend with high demands in the daytime and low demands in the night. The data profile in the third panel reflects a typical rural area, in which the traffic pattern is more or less random.}
\label{fig:typical_traffic_week}
\end{figure}

\subsection{GP-Based Wireless Traffic Modeling and Prediction}
In 5G, wireless traffic prediction is vital to resource allocation, load-aware management, and proactive control in C-RAN. In \cite{Xu19}, we proposed a distributed GP-based wireless traffic modeling and prediction framework that exploits the advanced C-RAN specifying the edge-based network infrastructure given in Section~\ref{sec:NetInf}. In the deployment area, several hundreds of micro base stations with fixed geographical positions are installed to serve mobile users and record the downlink physical resource block (PRB) usage (a wireless traffic usage indicator) versus time. In this work, the base stations serving as smart agents are first clustered into groups according to their geographical locations, and for each group an aggregated PRB usage prediction model is to be built. To this end, all the micro base stations in one cluster send their observed time series of PRB usage to an edge node, in which the data are aggregated, pre-processed and uniformly allocated to a number of parallel computing units. 

Specifically, a global GP regression model for the aggregated wireless traffic data of each cluster in the C-RAN is given as $y = f(t) + e$, where $y \in \mathbb{R}^{1}$ represents the PRB usage; $e$ is a Gaussian distributed noise term with zero mean and variance $\sigma_e^2$; $f(t)$ is a temporal GP as introduced in Eq.(\ref{eq:GP-model}) of Section~\ref{sec:lmodels}.

In comparison with the ``black-box'' deep learning models for sequential data modeling such as the recurrent neural network (RNN) and long-short term memory (LSTM), GP model owns better interpretability as prior information about the wireless traffic pattern can be encoded more easily into the kernel function design. As shown in Fig.~\ref{fig:typical_traffic_week}, the wireless traffic in our real data sets demonstrates the following general patterns: (1) \textit{weekly periodic pattern}, namely the variation in accordance with weekdays and weekends; (2) \textit{daily periodic pattern}, namely the variation in accordance with weekdays and weekends; and (3) \textit{deviations}, namely the small scale variation in addition to the above periodic trends. The first two patterns can be well captured by the periodic or the locally periodic kernel, while the third pattern can be well captured by the SE kernel or the Matern kernel. 

Our distributed GP for wireless traffic modeling and prediction falls in the FedLoc framework. Both the training and inference stages are performed in the edge nodes. Detailed workflow of model training is as follows. First, each base station in a specific cluster uploads its measured time series to the edge node. The aggregated data is then divided into $K$ portions by the edge node, and each portion is allocated to a local computing unit for distributed model training based on the cADMM introduced in Section~\ref{sec:fwp}.  The training framework achieves excellent tradeoff between the communication overhead and modeling accuracy, as explained in Section~\ref{sec:lmodels}. For each local computing unit, the required computational complexity can be reduced from $ \mathcal{O}(n^3) $ of the centralized, standard GP to $ \mathcal{O}(\frac{n^3}{K^3}) $, where $n$ is the number of the data points and $K$ the number of parallel computing units. 

In the online phase, one could use the generalized PoE \cite{Deisenroth15} to fuse the local predictions from all parallel computing units to approximate the global prediction. The generalized PoE model needs to introduce a set of fusion weight parameters, $\beta_i $, $i=1,2,...,K$, to take into account the importance of the local predictions. The resulting PoE predictive distribution is
\begin{equation}
\label{eq:gpoe-predictive}
p(f_*|\boldsymbol{x}_*, \mathcal{D}) \approx \prod_{i=1}^{K} p_i^{\beta_i}(f_*|\boldsymbol{x}_*, \mathcal{D}^{(i)}).
\end{equation}
The choice of $\beta_i $, $i=1,2,...,K$, is vital to the prediction. In \cite{Xu19}, we proposed to optimize the fusion weights according to the cross-validation criterion.  The corresponding weight optimization problem can be solved efficiently with convergence guarantee. More details about the optimization process can be found in \cite{Xu19}. 

In the above work, we considered a temporal GP for regression. Therein, each cluster of base stations is assumed to be independent other clusters. For enhanced prediction performance, we could use spatio-temporal GP that takes into account the correlations between different clusters. A straightforward way for building a spatio-temporal GP model is to introduce an extra kernel to account for the spatial correlations between different clusters and combine this spatial kernel with the aforementioned temporal kernel either through addition \cite{senanayake2016} or Kronecker product \cite{Bonilla08}. 

\begin{figure}[t]
	\includegraphics[width=0.48\textwidth,height=8cm]{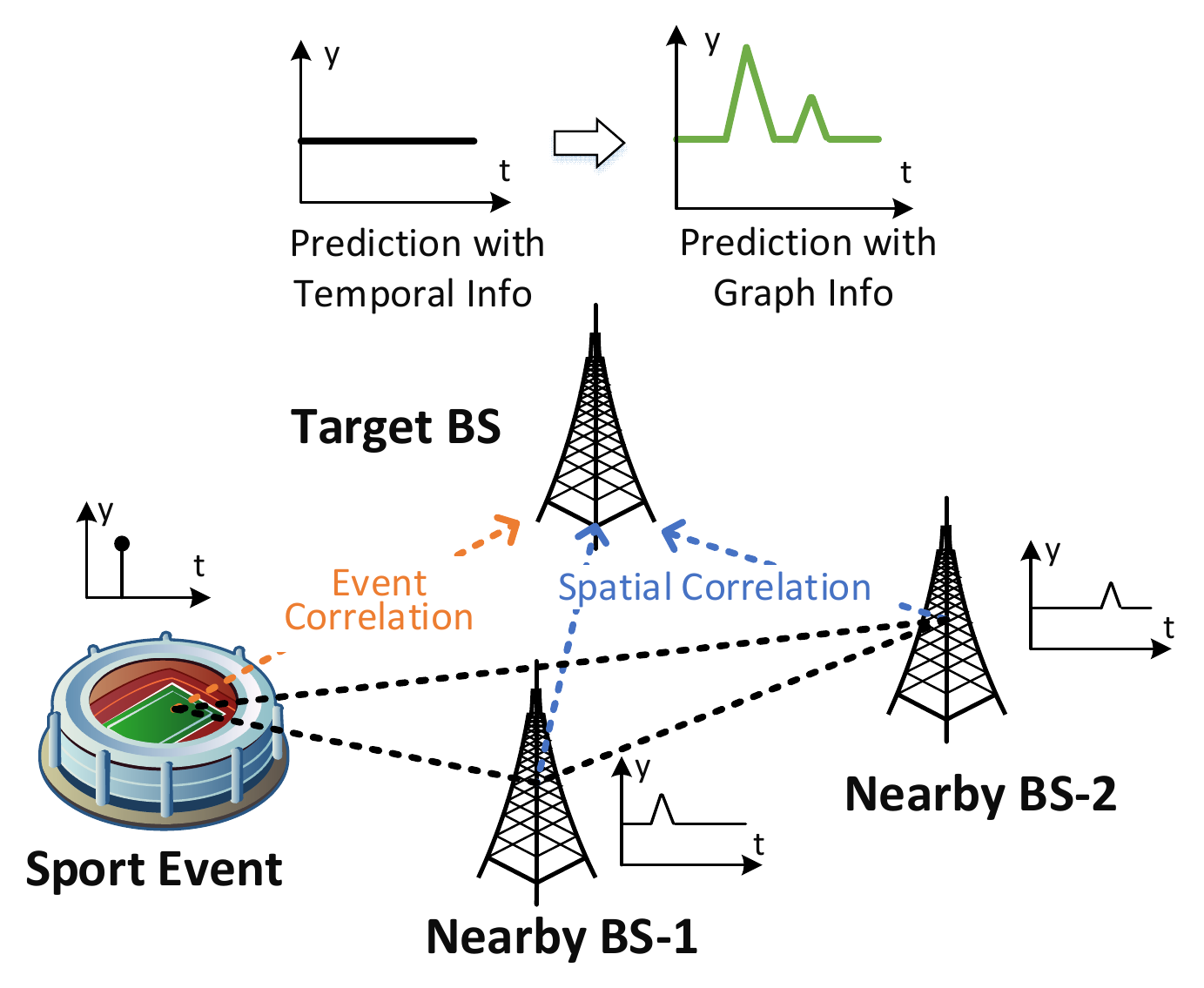}
	\caption{Conceptual illustration of graph GP for spatio-temporal data modeling. Both the spatial correlation and event correlation information are helpful to improve the prediction performance.}
	 \label{fig:ggp}
\end{figure}

The recently proposed graph GP provides another way for learning from high-dimensional data points living on non-Euclidean domains, see for instance \cite{vanderWilk17, Ng18, Walker19}. As such, graph GP allows for better non-local generalization thus can be used to model sophisticated correlation patterns across time and space. In the illustrating example in Fig.~\ref{fig:ggp}, a graph GP can be designed to capture three types of correlations, including: (1) temporal correlation as discussed above; and (2) spatial-temporal correlation, where closer geographical distance indicates higher correlation in the temporal observations, and (3) the event correlation, where an event nearby also indicates a higher probability of an abrupt traffic change. It is noteworthy that graph GP is still under development where many directions remain to be explored, e.g., kernel design, stability issue, and distributed processing among others.

\subsection{Other Potential Use Cases}
Due to space limitations, we are unable to give a full list of all FedLoc related use cases with details. However, we want to briefly demonstrate the following three use cases due to their ever-increasing popularity. 

(1) Radio feature map construction. The proposed FedLoc framework can be used by a number of collaborating mobile users to build accurate radio feature maps, such as RSS map and magnetic field map, for indoor venues. In \cite{Yin17}, we proposed a distributed, recursive GP framework for building indoor RSS maps. Therein, a batch of mobile users was employed to collect RSS measurements from a dozen of WiFi access points at Ericsson research, Linkoping, Sweden. In the training phase, each mobile user trains a local GP empowered RSS map individually, while in the inference phase a global prediction is obtained by fusing all the local GP models via the classical Bayesian committee machine. A follow-up work was then proposed in \cite{Zhao18}. These works can be revised to fit a global GP model in the training phase using the ADMM-based GP hyper-parameter optimization algorithm introduced in Section~\ref{sec:fwp}.

(2) Simultaneous localization and mapping (SLAM) for three-dimensional (3D) indoor scenario construction. The proposed FedLoc framework can also be used for a number of collaborating robots or low-flying unmanned aerial vehicles (UAVs) equipped with cameras and LIDAR to reconstruct a 3D indoor scenario. 
A generic SLAM model \cite{Gustafsson2010} is given as follows:
\begin{align}
\boldsymbol{x}_t & = f(\boldsymbol{x}_{t-1}, \boldsymbol{u}_{t-1}) + \boldsymbol{e}_{t-1}, \nonumber \\
\boldsymbol{m}_{t} &= \boldsymbol{m}_{t-1}, \nonumber \\
\boldsymbol{y}_t & = g(\boldsymbol{x}_t, \boldsymbol{m}_{t}, \boldsymbol{u}_{t}) + \boldsymbol{n}_t, 
\label{eq:slam}
\end{align}
where the dynamic motion model takes an additional inertial input $\boldsymbol{u}_t$ of the sensory data from odometer, accelerometer, gyroscope, and there is an additional map memory state, $\boldsymbol{m}_{t}$, in which the positions of the landmarks are updated and stored. We could potentially modify the GPSSM framework for the federated SLAM. Different from the use cases given in Section~\ref{sec:use-cases}, federated SLAM imposes more stringent requirements on both the computational power of the mobile devices and the data throughput of the network, when dealing with 3D environment reconstruction. The commercial 5G network and futuristic wide-band generations (B5G and 6G) could make the federated SLAM possible. Some recent attempt in this regard can be found in \cite{LiWang19}. 

\begin{figure}
\centering
\includegraphics[width=0.47\textwidth]{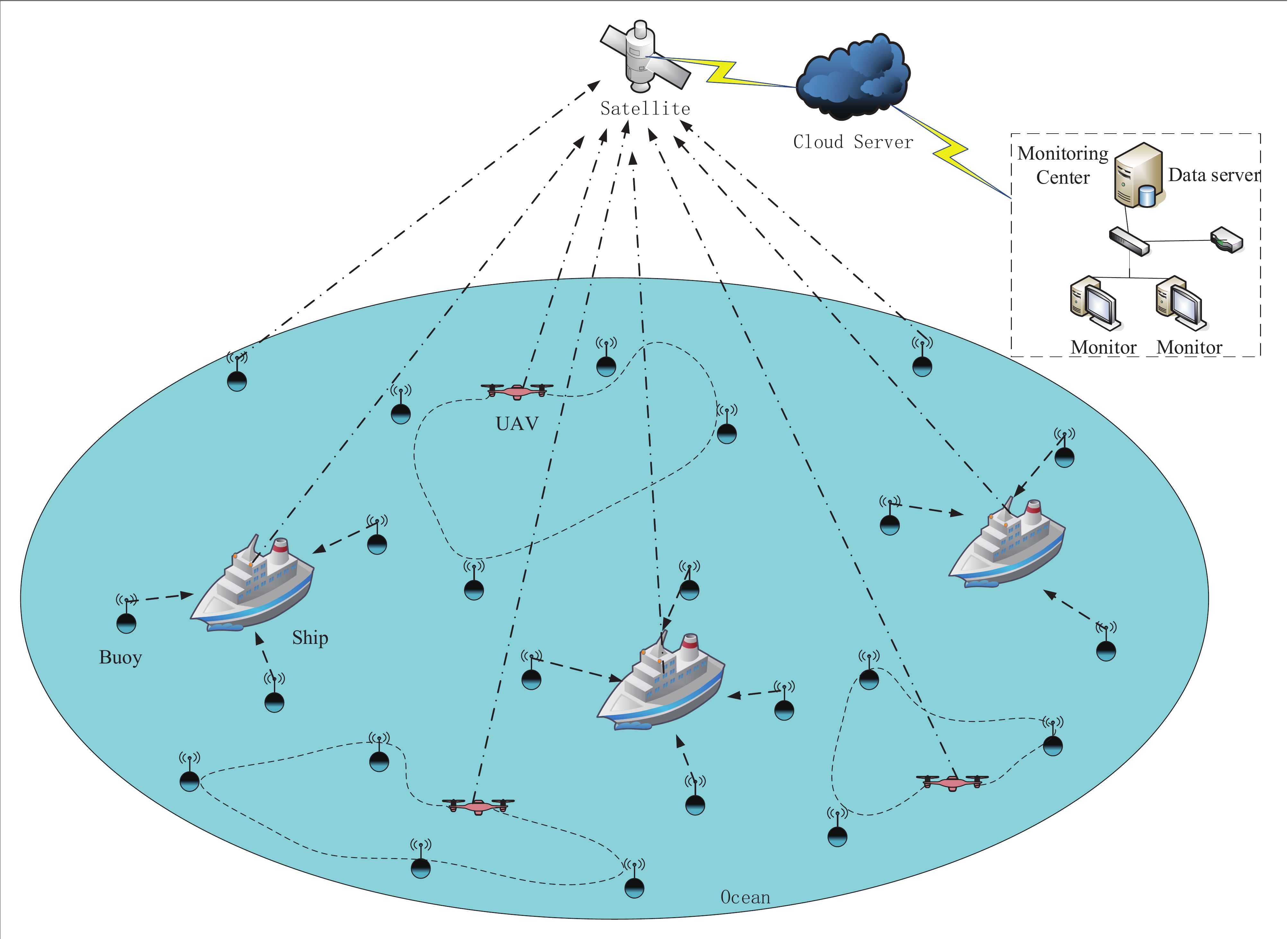} 
\caption{A conceptual picture of Ocean-of-Things. Here, buoys can be seen as smart agents, ships and UAVs as edge nodes, and satellite as central node with cloud facility.} 
\label{fig:OoT} 
\end{figure}

(3) Ocean-of-Things (OoT) \cite{Waterston19}. So far we have solely considered ground applications. In addition, there will be a plethora of emerging OoT applications that can benefit from our FedLoc framework. We show a conceptual picture of OoT in Fig.~\ref{fig:OoT}, where the whole network comprises a large number of spatially distributed buoys, some moving ships and UAVs, and satellites. The buoys are analogous to micro base stations on the ground, serving as smart agents, and they can perform data collection and monitor local environment. New fashioned buoys will be equipped with different sensors, ranging devices, GPS, and low-profile AI chips. They can be used to measure the ocean surface temperature, sea state, sound speed, etc, and track multi-target trajectories. The measured local data can be uploaded either to a moving ship or a moving UAV, which can be regarded as edge node. In addition to information transmission, the UAVs can also be used to charge the buoys if they are wireless powered \cite{Chen2018uav, Hu2019optimal}. Each edge node maintains a local update of the learning model for spatio-temporal data processing and transmits the hyper-parameter estimate to the satellite cloud for consensus. In contrast to the ground IoT applications, the buoys may have insufficient on-board processing capability and relatively short communication range compared with a micro base station. However, the communication channels on the sea are mostly in line-of-sight. Since the buoys may be owned by different operators, privacy-preservation can not be ignored either. 

\section{Results}
\label{sec:results}
In this section, we show the effectiveness of the FedLoc framework with two examples evaluated using real data sets. In the first example, we adopt GP as the learning model and mainly focus on the effectiveness of the distributed training of a small batch of model hyper-parameters. In the second example, we adopt DNN as the learning model and focus on practical implementation aspects.

\subsection{GPSSM for Indoor Target Tracking}
In this section, we will demonstrate the first example of applying the FedLoc framework for target tracking. The experimental setup aims for a quick and practical deployment of the framework, thus may not be theoretically optimal. Our focus is on both the training and prediction performance of the global, centralized model versus its distributed approximation under the FedLoc framework.

Due to space limitations, we will only show some results for the transition function in GPSSM. The model is $\boldsymbol{x}_{t+1} = f(\boldsymbol{x}_{t}) + \boldsymbol{e}_{t}$, where the vector $\boldsymbol{x}_t = [x_t, y_t]^T$ contains the 2-D position of a pedestrian at time instance $t$. We apply individual GPs for each dimension, namely, we let 
\begin{subequations}
\begin{align}
x_{t+1} &= f_{x}(\boldsymbol{x}_{t}) + \boldsymbol{e}_{x, t}, \\ 
y_{t+1} &= f_{y}(\boldsymbol{x}_{t}) + \boldsymbol{e}_{y, t},
\end{align}
\end{subequations}
where both $f_{x}(\boldsymbol{x}_{t})$ and $f_{y}(\boldsymbol{x}_{t})$ are modeled by GP; for instance, we let 
\begin{equation}
f_{x}(\boldsymbol{x}_{t}) \sim GP(m_{x}(\boldsymbol{x}_{t}), k_{x}(\boldsymbol{x}_{t}, \boldsymbol{x}_{t'})).
\end{equation}
For clear exposition, we let the mean function $m_{x}(\boldsymbol{x}_{t})$ be zero and the kernel function $k_{x}(\boldsymbol{x}_{t}, \boldsymbol{x}_{t'})$ be the ARD kernel, i.e.,
\begin{equation}
k_{x}(\boldsymbol{x}_{t}, \boldsymbol{x}_{t'}) \!\!=\!\! \sigma_{s,x}^2 \exp \!\left[ -\frac{(x_t - x_{t'})^2}{l_{xx}}- \frac{(y_t -y_{t'})^2}{l_{xy}} \right],
\end{equation}
where the kernel hyper-parameters are $[\sigma_{s,x}^2, l_{xx}, l_{xy}]^T$. For the $y$-dimension, we adopt a similar ARD kernel, $k_{y}(\boldsymbol{x}_{t}, \boldsymbol{x}_{t'})$, but with a different set of kernel hyper-parameters $[\sigma_{s,y}^2, l_{yx}, l_{yy}]^T$. 

The above GP models can be trained globally with a training data set $\mathcal{D}$ via the global, centralized maximum-likelihood estimation shown in Eq.(\ref{eq:GP-global-obj}). We know from Section~\ref{sec:lmodels} that the computational complexity scales as $\mathcal{O}(n^3)$ for centralized model training. Using the FedLoc framework is beneficial. On the one hand, mobile users can collect their own local training data without worrying about the data leakage issue, which may effectively encourage more people to collaborate. By adopting the cADMM or the pxADMM introduced in Section~\ref{sec:fwp} to approximate the global model hyper-parameters in a distributed manner, the overall computational complexity can be reduced to $\mathcal{O}(n^3/K^3)$, where $K$ is the number of the collaborating mobile users. This work can be seen as a collaborative, data-driven method for learning the human walking trajectory, which is valuable for us to understand the behavior of pedestrians and predict their future positions. 
%

To evaluate the performance of the FedLoc, we collected a data set in a live indoor office environment, as was shown in Fig.~\ref{fig:figure3}. This data set contains more than 50 trajectories with around 25,000 samples. In the training phase, three mobile users each collected 15 trajectories. Each mobile user obtained an approximation of the global GP model shown in Eq.(\ref{eq:GP-global-obj}) using its local 15 trajectories. In the test phase, we use the model hyper-parameters trained from the FedLoc to perform posterior prediction of the next state given a novel current state. 

We compare two distributed GP hyper-parameter optimization schemes: (1) pxADMM-GP with the regularization parameters $\rho_i = 500$ and $L_i = 5000$, $\forall i$; and (2) cADMM-GP with $\rho_i = 500$, for $i=1,2,3$. We set the values for $\rho_i$ and $L_i$ empirically. We consider convergence when the difference in all optimization variables between two consequent iterations is within $10^{-3}$. The computer program was implemented using MATLAB and executed on an ordinary computer with 4 cores. 

\begin{figure}[t]
	\centering
	\subfloat[]{\includegraphics[width=0.45\columnwidth]{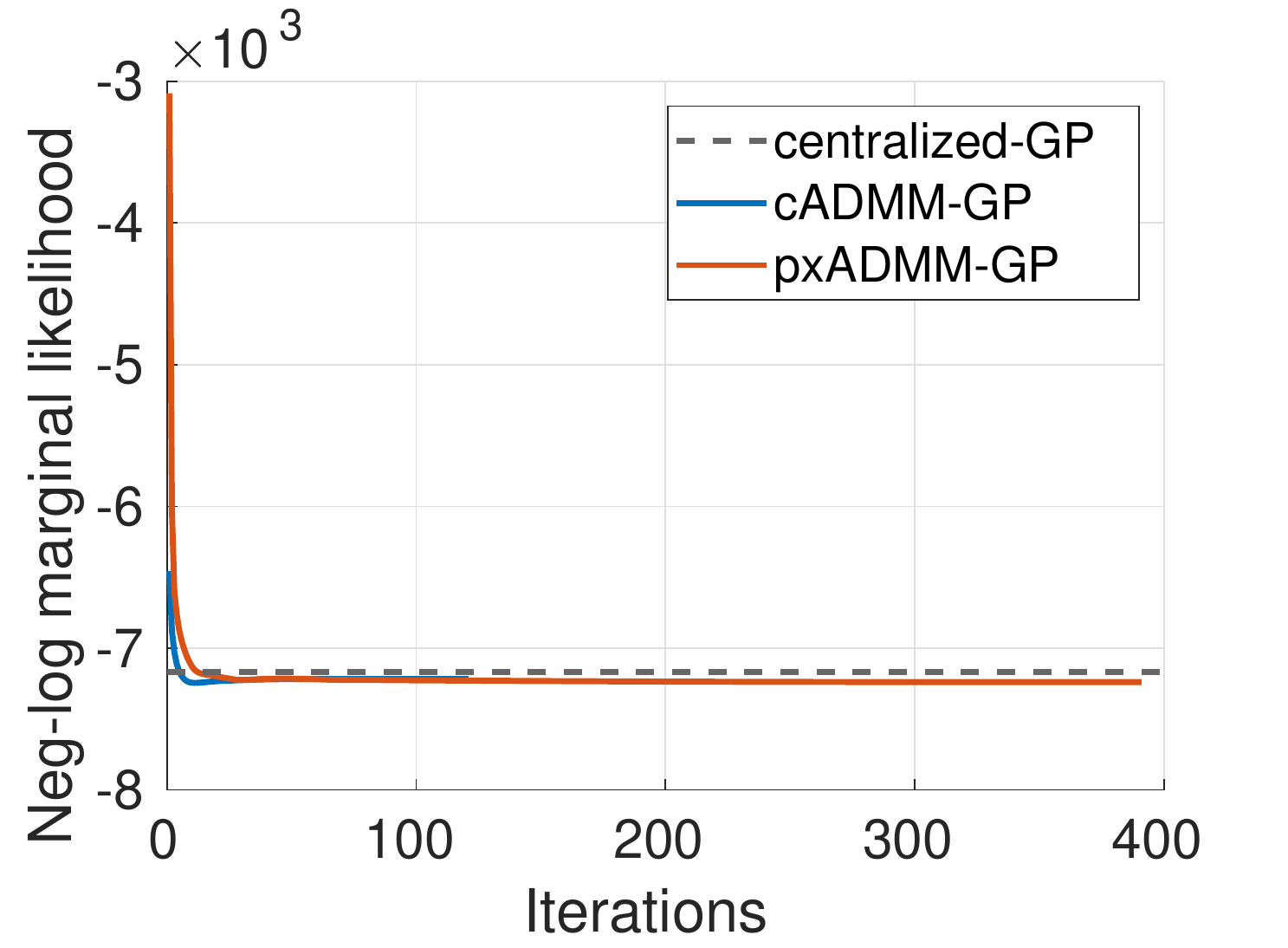}}
	\hfil
	\subfloat[]{\includegraphics[width=0.45\columnwidth]{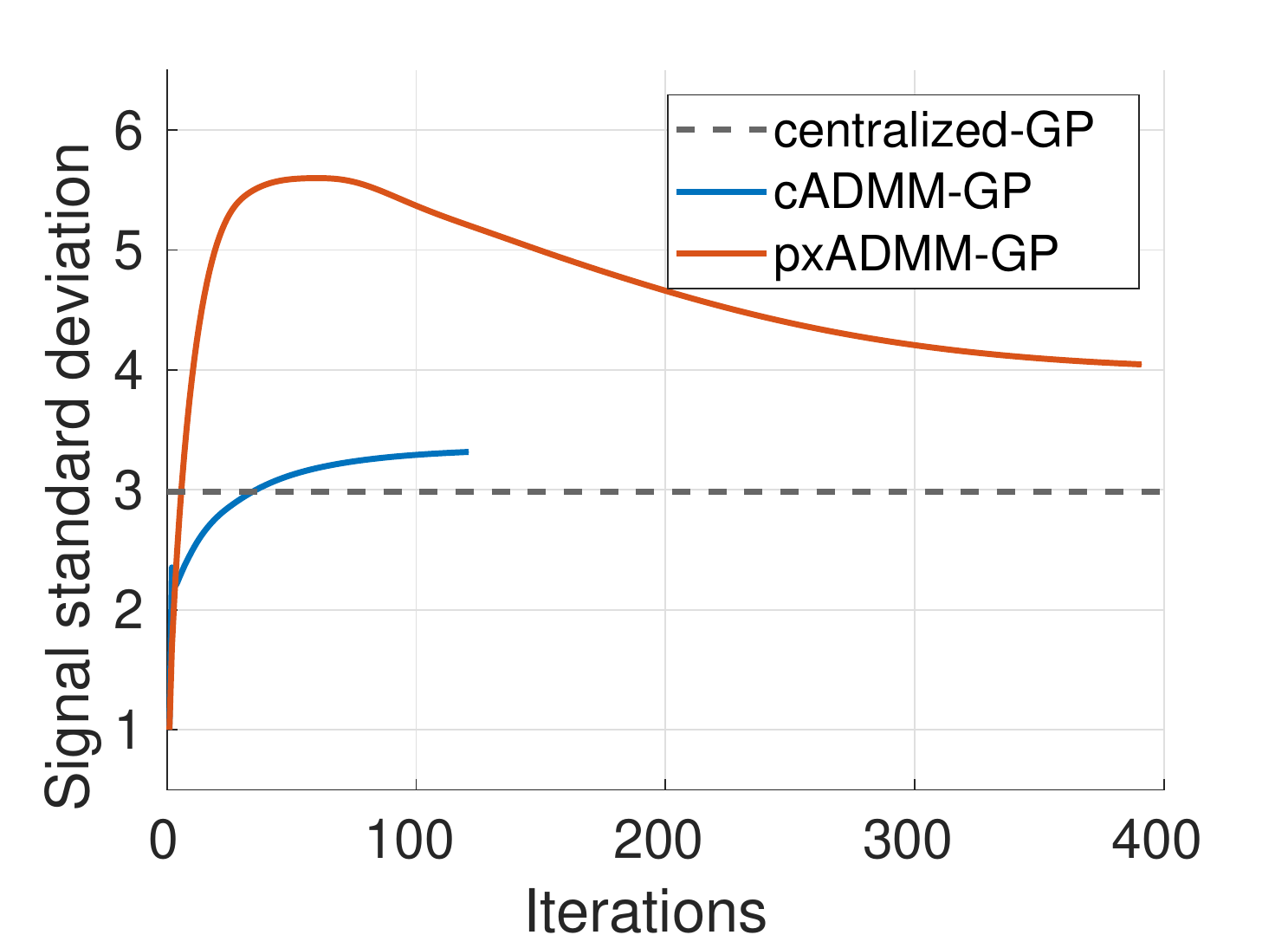}}
	
	\subfloat[]{\includegraphics[width=0.45\columnwidth]{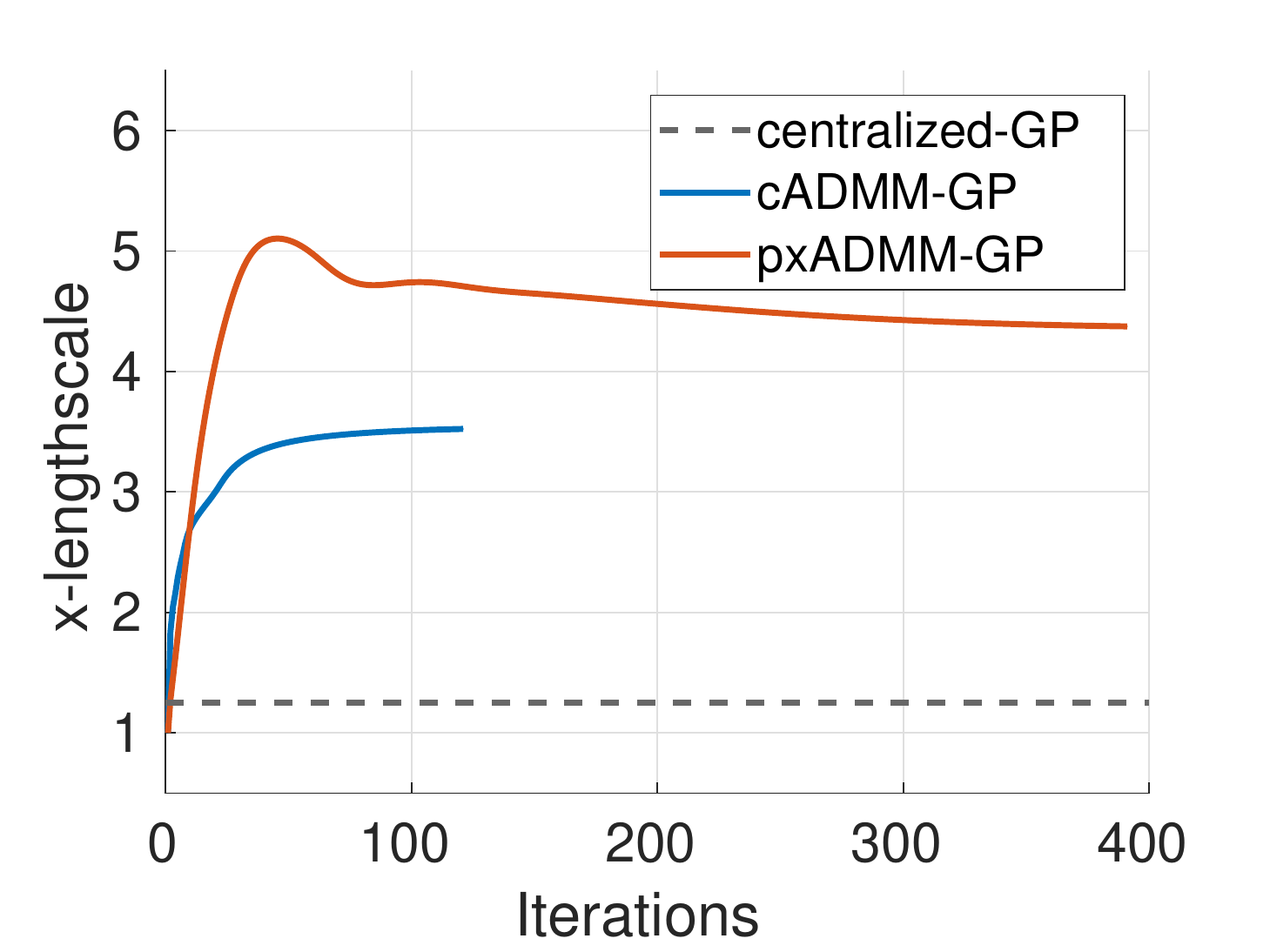}}
	\hfil
	\subfloat[]{\includegraphics[width=0.45\columnwidth]{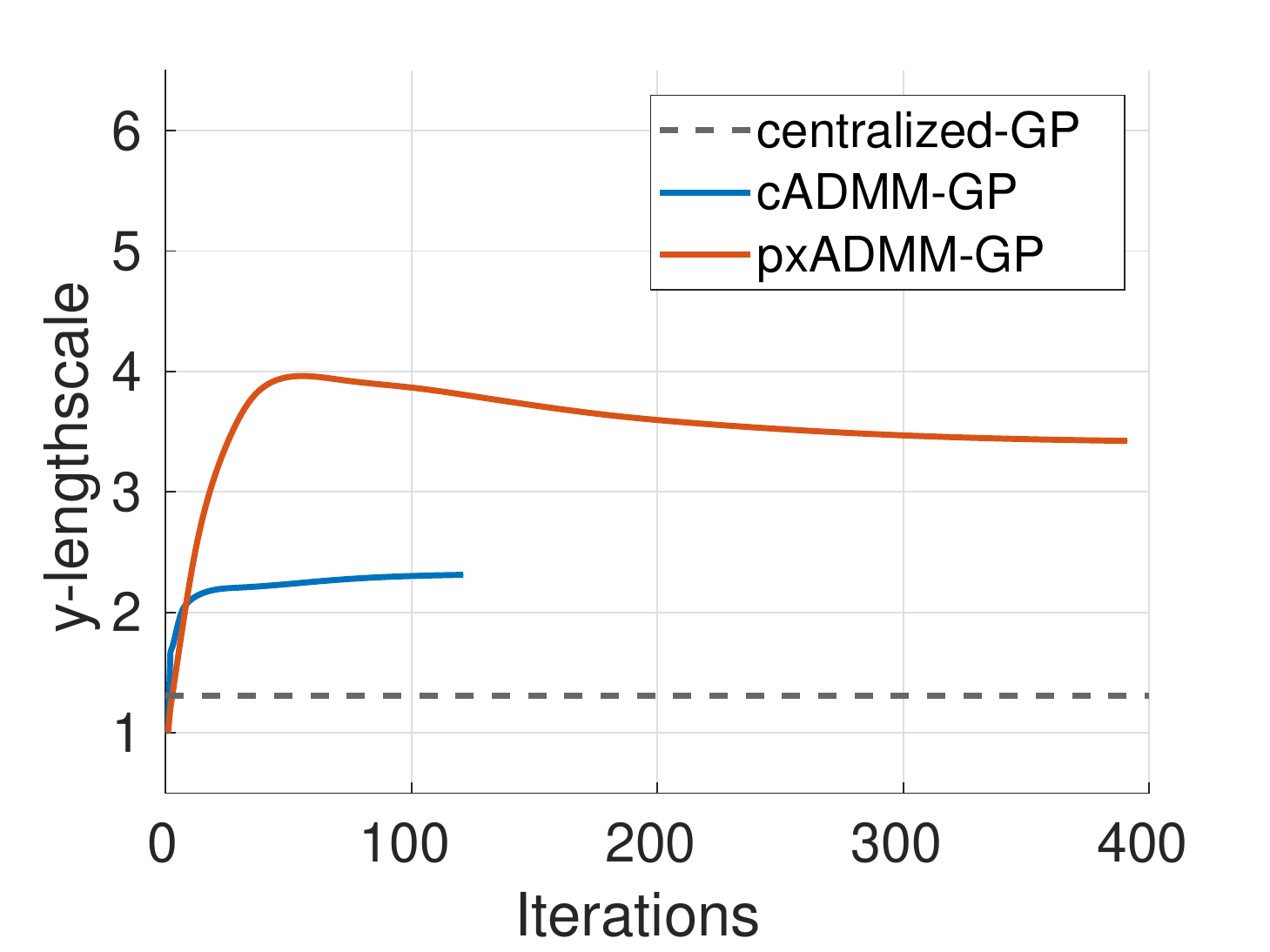}}
	\caption{For GP modeling along the $x$-dimension, we show the negative log-marginal likelihood functions (centralized formulation refer to Eq.(\ref{eq:GP-global-obj}) and distributed formulation refer to Eq.(\ref{eq:GP-local-approx})) in sub-figure (a); and the ARD kernel hyper-parameter estimates as a function of training iterations for the $3$ input variables using pxADMM-GP and cADMM-GP in sub-figures (b-d) for model variance, length-scale in $x$, and length-scale in $y$, respectively.}
	\label{fig:results-x-dimension}
\end{figure}

\begin{figure}[t]
	\centering
	\subfloat[]{\includegraphics[width=0.45\columnwidth]{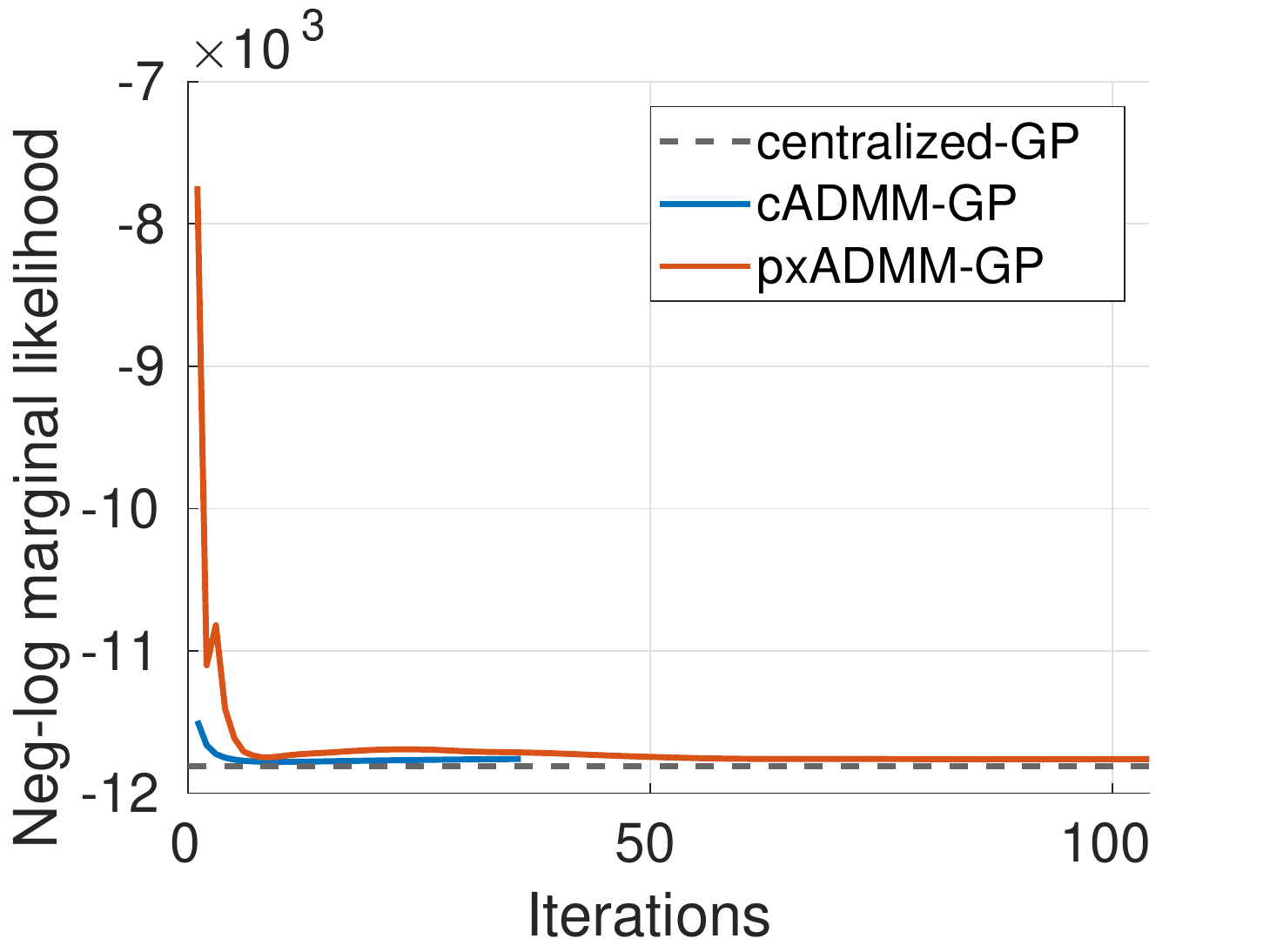}}
	\hfil
	\subfloat[]{\includegraphics[width=0.45\columnwidth]{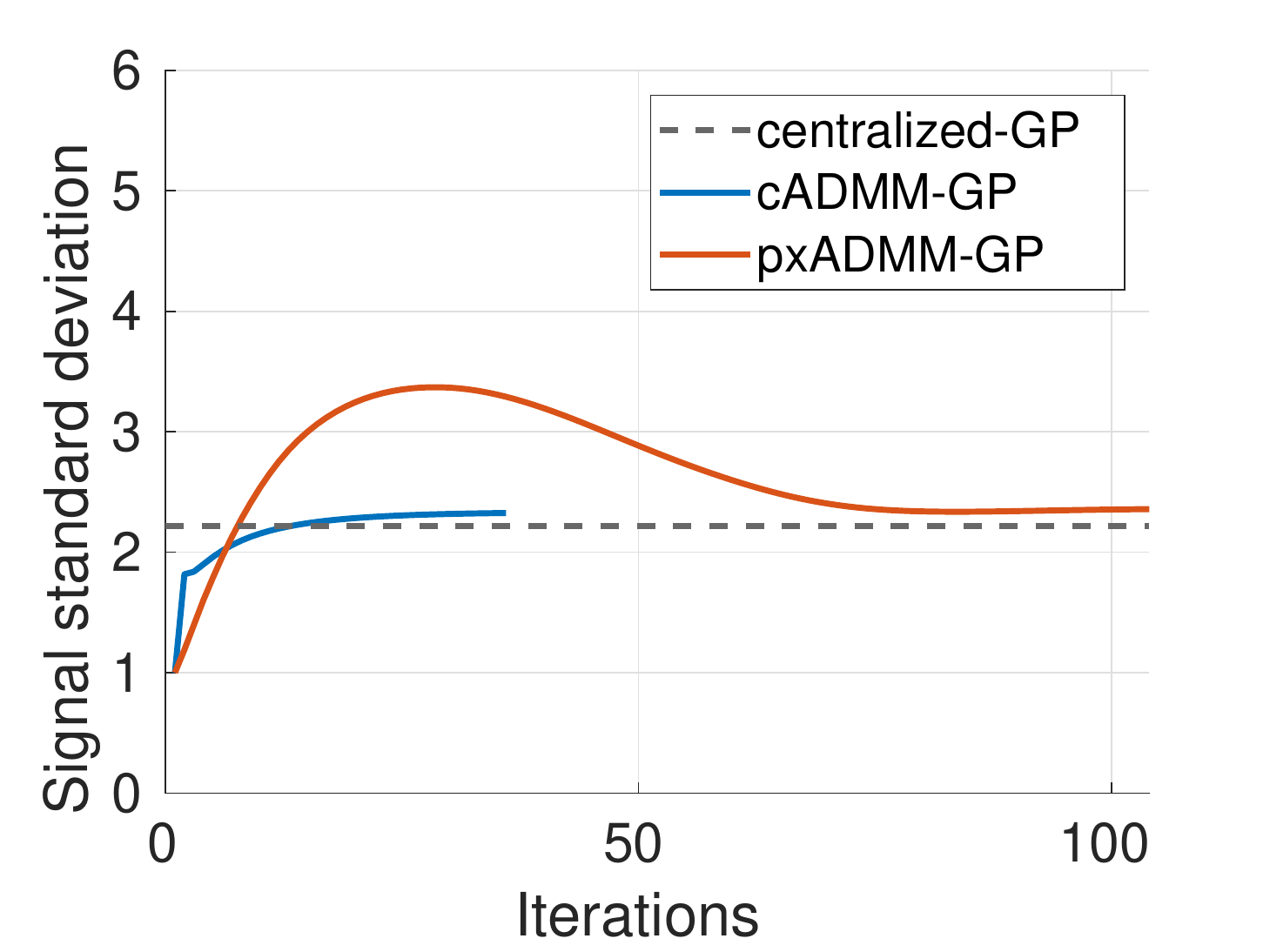}}
	
	\subfloat[]{\includegraphics[width=0.45\columnwidth]{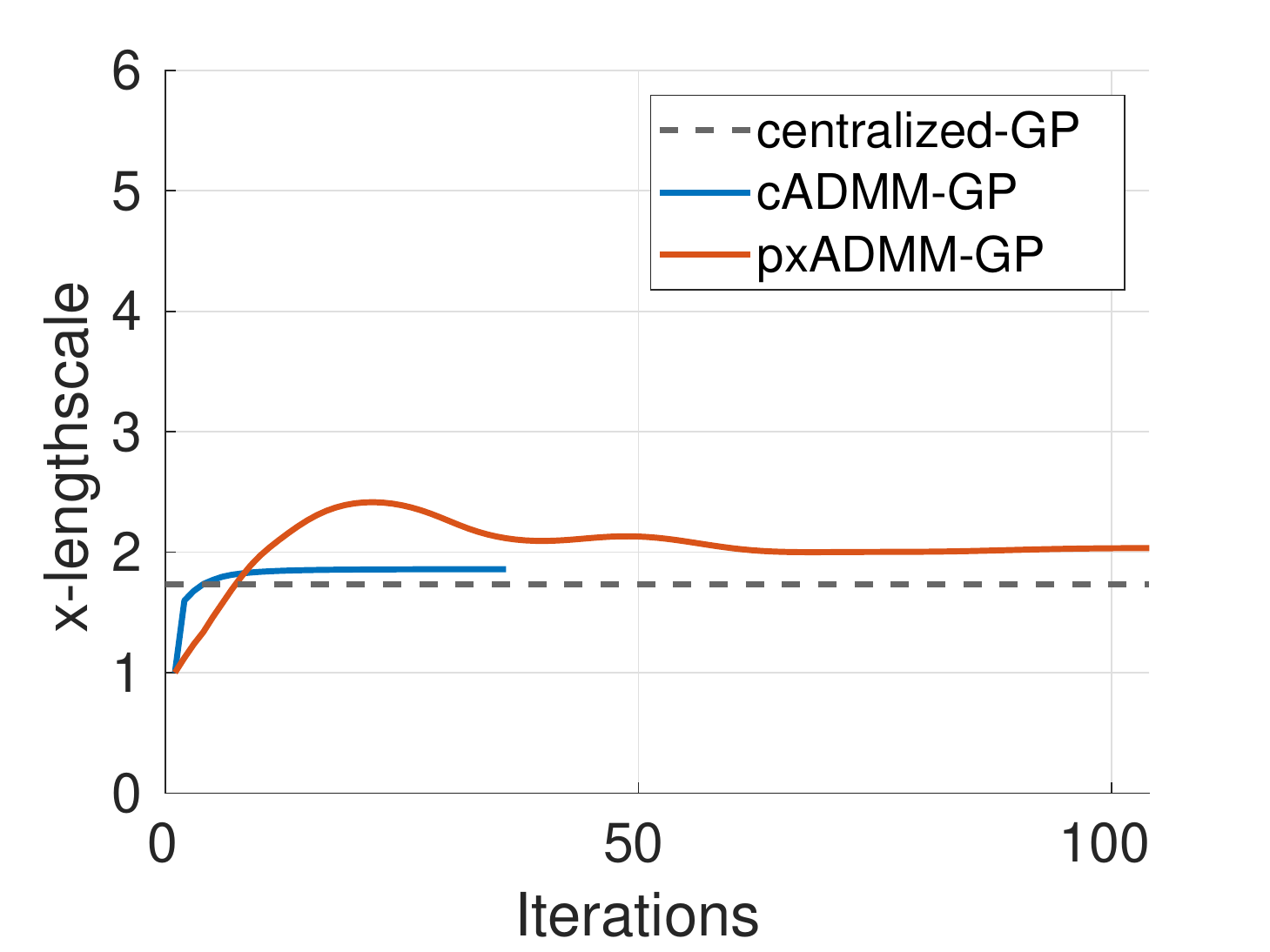}}
	\hfil
	\subfloat[]{\includegraphics[width=0.45\columnwidth]{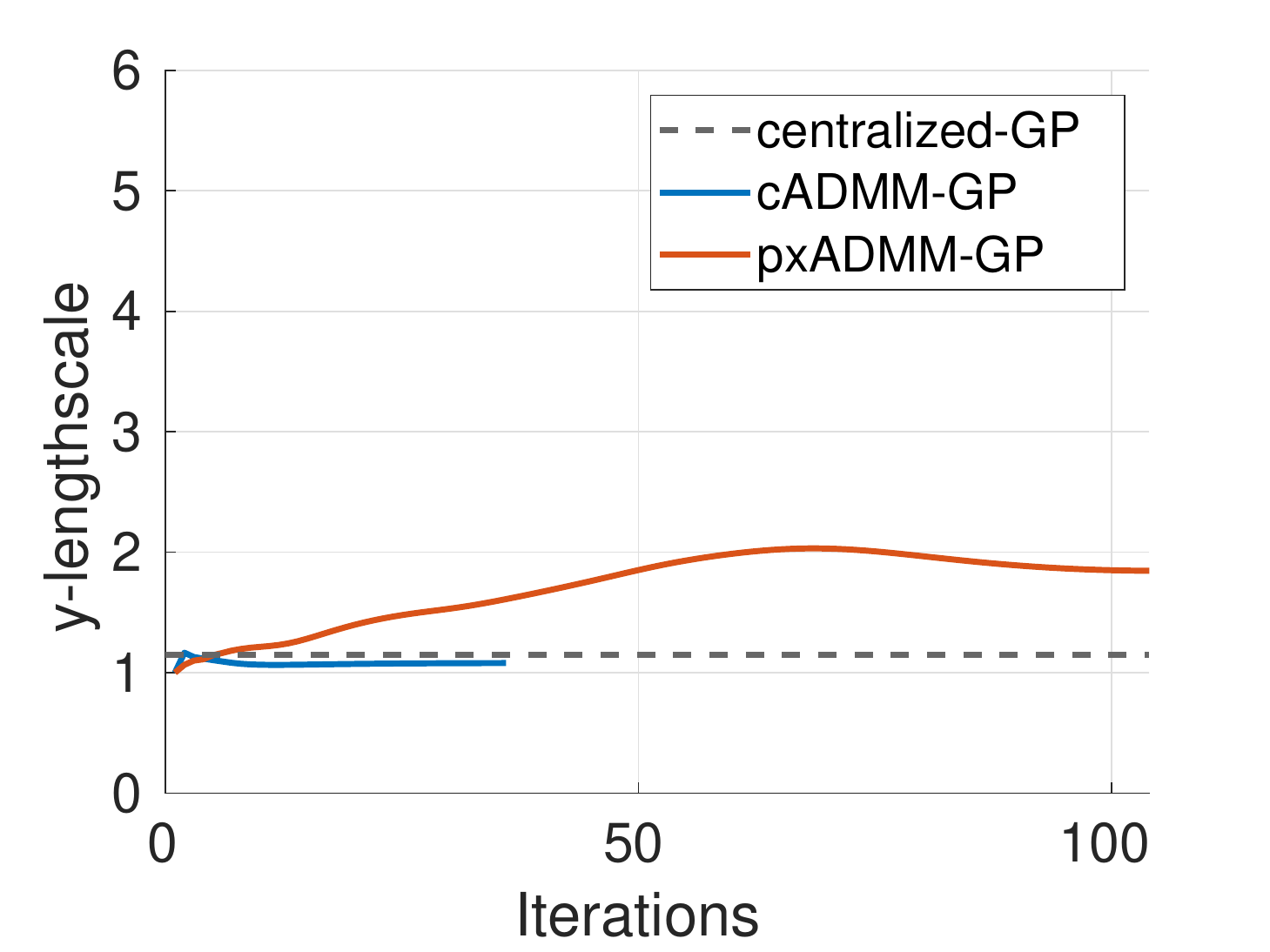}}
	\caption{Convergence results for the GP modeling along the $y$-dimension.}
	\label{fig:results-y-dimension}
\end{figure}

We show the model training results for both dimensions ($x$ and $y$) in Fig.~\ref{fig:results-x-dimension} and Fig.~\ref{fig:results-y-dimension}. The distributed schemes converge to different model hyper-parameter estimates compared with the ones trained centrally for the global model. One reason is that the distributed scheme uses a different cost function as shown in Eq.(\ref{eq:GP-local-approx}), which corresponds to approximating the kernel matrix $K(\boldsymbol{X},\boldsymbol{X}; \boldsymbol{\theta})$ to a block diagonal matrix. Despite the difference in the hyper-parameter estimates, the corresponding negative log-marginal likelihood as well as the overall prediction root-mean-squared-error (RMSE in meters) are fairly close. 
From the computational time (CT) shown in Table~\ref{tab:table2}, we observed that the pxADMM-GP scheme consumed the least computational time. On one hand, the pxADMM-GP scheme circumvents frequent gradient synchronizations and used less iterations toward convergence than the cADMM-GP scheme. On the other hand, the closed-form proximal update w.r.t.\ the local hyper-parameters only requires to compute the expensive matrix inversion once. 
\begin{table}
	\caption{Comparisons of two distributed GP model training schemes.}
	\label{table:RMSE}
	\centering
	\begin{tabular}{ccc}
		\hline
		  & pxADMM-GP  & cADMM-GP \\ 
		\hline
		RMSE & 0.1368m & 0.1353m \\
		\hline
		CT & 714s & 10838s \\
		\hline
	\end{tabular}
	\label{tab:table2}
\end{table}

\subsection{Outdoor Vehicle Navigation with Low-Sampling-Rate GPS}
In this section, we will demonstrate the application of FedLoc with DNN models for smart vehicle navigation using low-sampling-rate GPS signals, which was introduced as a representative use case in Section~\ref{sec:use-cases}. 

\begin{figure}
\centering
\includegraphics[width=0.4\textwidth]{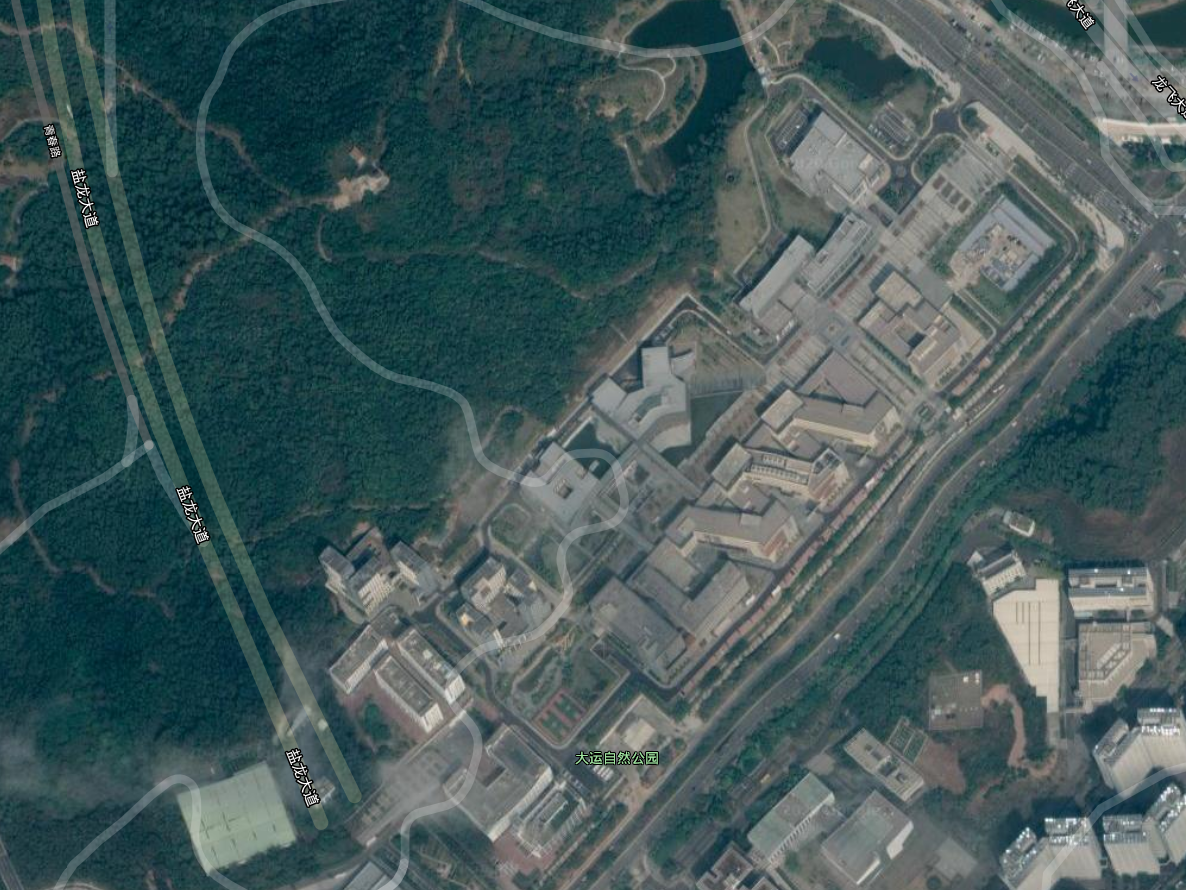} 
\caption{The satellite map of the CUHK(SZ), where we collected the outdoor vehicle navigation data along two different routes.} 
\label{fig:cuhksz-map} 
\end{figure}

We start by introducing the implementation setups of our new proposed federated learning empowered navigation system prototype. First, real data sets (for both training and test) were collected by three collaborating users with their own private car driving on the campus of The Chinese University of Hong Kong (Shenzhen), see Fig.~\ref{fig:cuhksz-map}. During the data collection process, each car was equipped with a smartphone (Xiaomi), facing upwards and heading to the moving direction of the car. The sensor data were uploaded to the server through WiFi on the fly. These three collaborating users traveled around the campus and collected various trajectories of smartphone sensory data that contain real-time motion information of their vehicles. The duration of each trajectory ranges from a few minutes to dozens of minutes. 

After collecting all training data sets, we adopted the FedLoc framework to train the two DNNs as was introduced in Section~\ref{sec:use-cases} for calibrating the sensor data, one for the velocity and the other one for the yaw angle, so that accurate navigation can be obtained even with low-sampling-rate GPS signals. Two DNNs with five hidden layers (3000-3000-2000-1000-500) are selected as the global model in our prototype, which can be replaced with more sophisticated models, such as the LSTM, for high-dimensional time series. The input is the sensor data measured in a specific time window with dimension 600 for the first DNN or with dimension 401 for the second DNN, while the output is a scalar. In the training phase, the global model is updated by the three collaborating users according to Algorithm~1. Specifically, we tried two different model training algorithms, namely the FedAvg algorithm and the FedProx algorithm introduced in Section~\ref{sec:fwp}. We set the learning rate to $10^{-4}$ for both the FedAvg and FedProx algorithms. For the FedProx algorithm, the additional regularization parameter is set to $10^{4}$. In the following, we consider two different experimental setups to mimic near i.i.d. and balanced data across the users as well as non-i.i.d. and unbalanced data across the users, to test the FedLoc framework. We elaborate on the two different setups in Table~\ref{tab:setups}.
\begin{table}
\caption{Two different experimental setups.}
\label{tab:setups}
\centering
\begin{tabular}{|c|c|c|c|c|}
\hline
\multicolumn{2}{|c|}{}                                                                                 & \textbf{user 1} & \textbf{user 2} & \textbf{user 3} \\ \hline
\multirow{2}{*}{\textbf{\begin{tabular}[c]{@{}c@{}}i.i.d. \\   \& balanced data \end{tabular}}}        & \textbf{route 1} & 4               & 4               & 4               \\ \cline{2-5} 
                                                                                        & \textbf{route 2} & 0               & 0               & 0               \\ \hline
\multirow{2}{*}{\textbf{\begin{tabular}[c]{@{}c@{}}non-i.i.d.\\   \& imbalanced data \end{tabular}}} & \textbf{route 1} & 0               & 2               & 6               \\ \cline{2-5} 
                                                                                        & \textbf{route 2} & 2               & 0               & 1               \\ \hline
\end{tabular}
\end{table}

\begin{figure}[htbp]
\centering
\subfloat[]{\includegraphics[width=8cm]{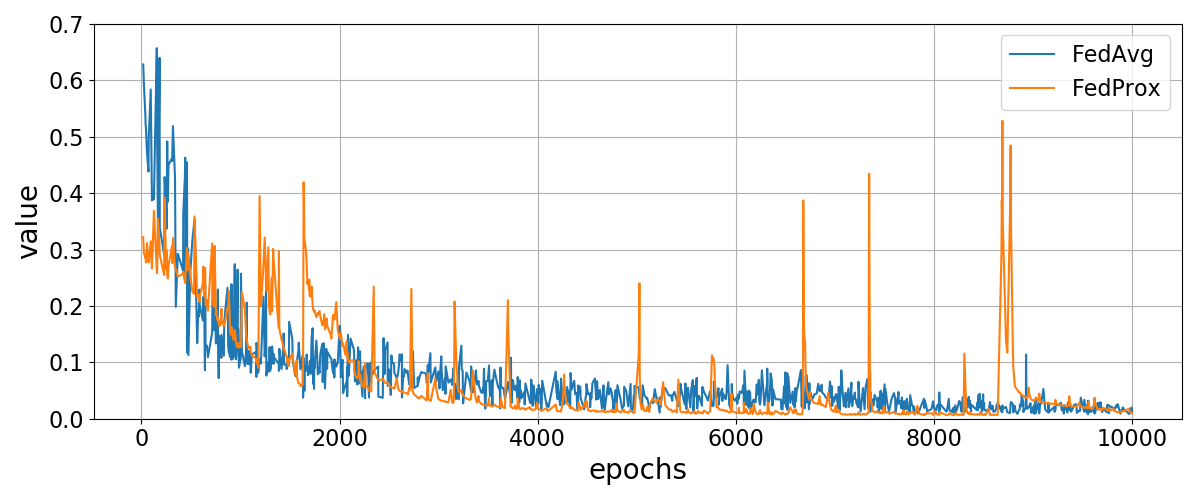}}
\hfil
\subfloat[]{\includegraphics[width=8cm]{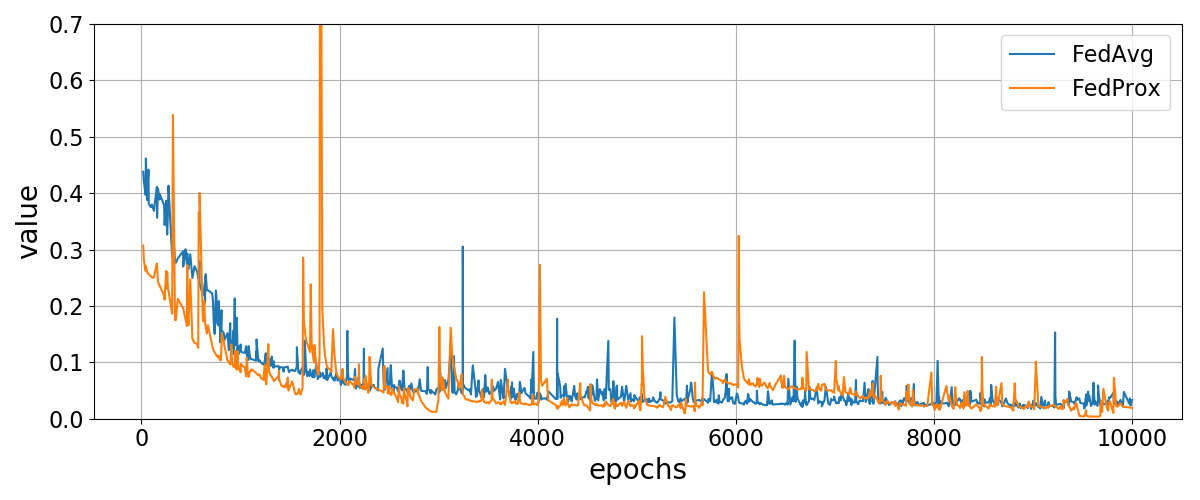}}
\caption{Training loss versus optimization epochs for two different optimization algorithms. (a) Setup 1: near i.i.d. and balanced data; and (b) Setup 2: non-i.i.d. and imbalanced data.}
\label{fig:training-loss-fedavg-fedprox}
\end{figure}

We show the training performance of both the FedAvg and FedProx algorithms in Fig.~\ref {fig:training-loss-fedavg-fedprox}. Both algorithms can achieve a low training loss after a certain number of epochs. In our experiments, the FedProx algorithm unfortunately did not demonstrate smoother and more stable convergence profile than that of the FedAvg algorithm. The reason may lie in the improper setting of the regularization parameter of the FedProx algorithm, which is supposed to help achieve good trade off between the training loss and the discrepancy between the global model and local ones.

\begin{figure}[htbp]
	\centering
	\subfloat[]{\includegraphics[width=4.2cm]{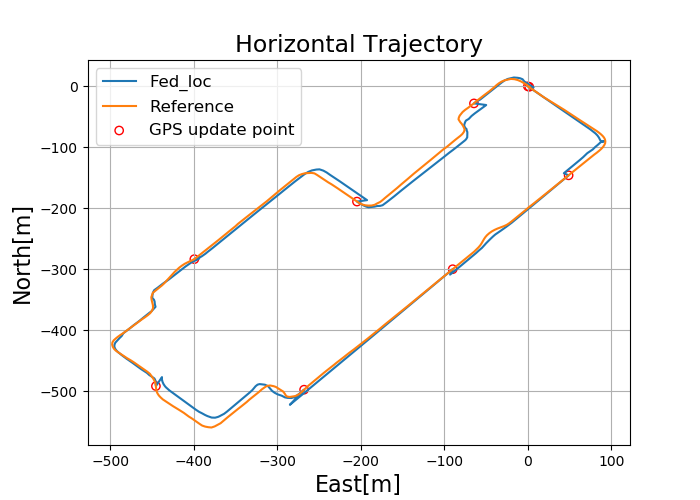}}
    \hfil 
    \subfloat[]{\includegraphics[width=4.2cm]{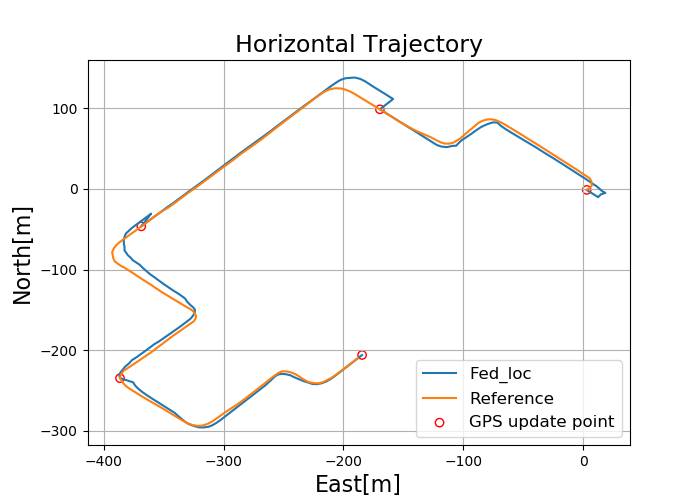}}
    
   	\subfloat[]{\includegraphics[width=4.2cm]{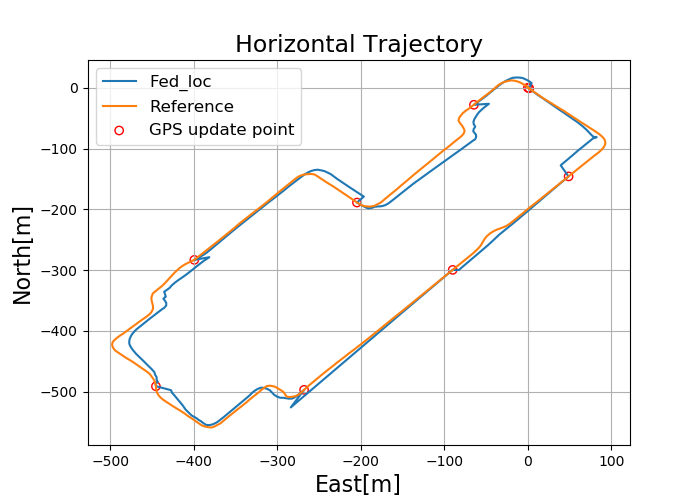}}
    \hfil 
    \subfloat[]{\includegraphics[width=4.2cm]{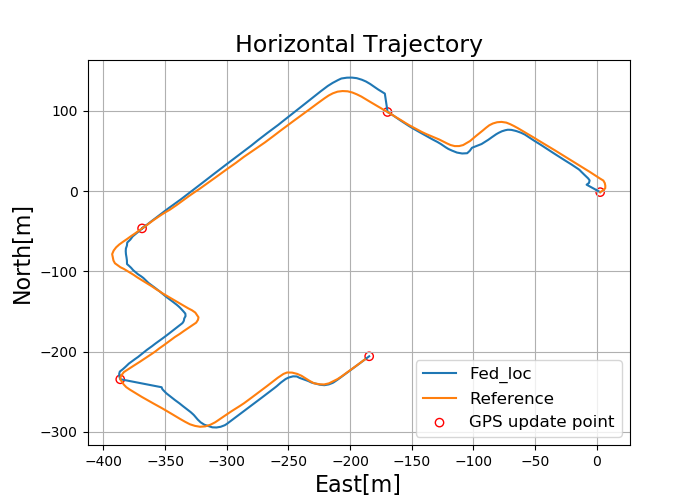}}
	\caption{The test performance on two test trajectories provided by the two algorithms for i.i.d. and balanced data setup. Subfigures (a) and (b) are drawn for the FedAvg algorithm; Subfigures (c) and (d) are drawn for the FedProx algorithm.}
\label{fig:test-performance-fedavg-fedprox-iid}
\end{figure}

\begin{figure}[htbp]
	\centering
	\subfloat[]{\includegraphics[width=4.2cm]{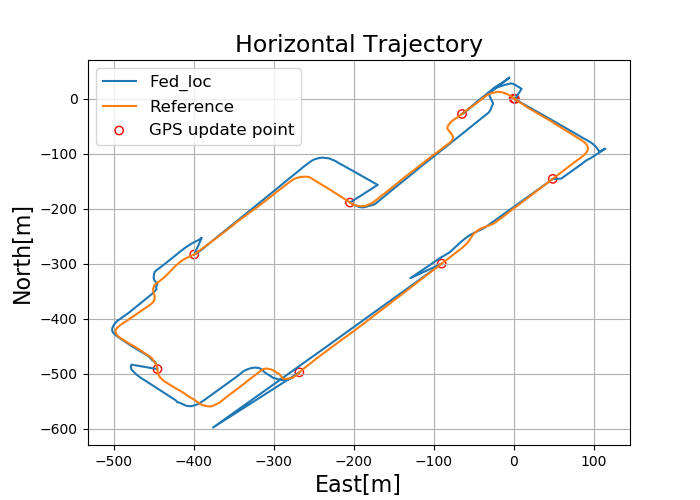}}
    \hfil 
    \subfloat[]{\includegraphics[width=4.2cm]{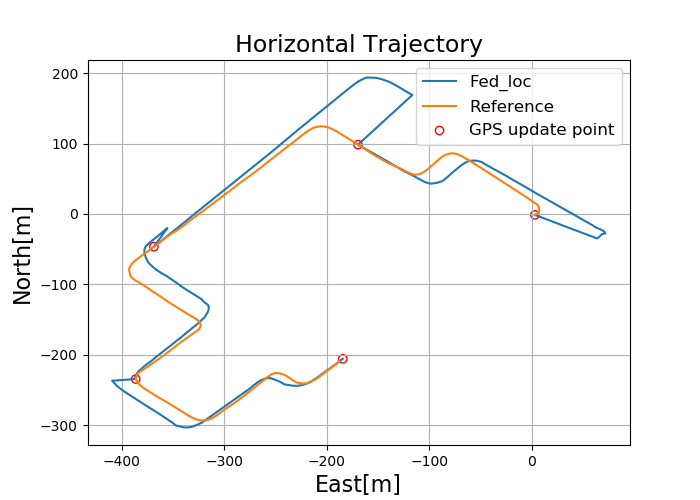}}
    
    \subfloat[]{\includegraphics[width=4.2cm]{iid-fedprox-61-test-traj}}
    \hfil 
    \subfloat[]{\includegraphics[width=4.2cm]{iid-fedprox-62-test-traj}}
	\caption{The test performance on two test trajectories provided by the two algorithms for non-i.i.d. and imbalanced data setup. Subfigures (a) and (b) are drawn for the FedAvg algorithm; Subfigures (c) and (d) are drawn for the FedProx algorithm.}
\label{fig:test-performance-fedavg-fedprox-non-iid}
\end{figure}

Lastly, we test the trained global learning model with two new trajectories of route 1. The GPS reference signals are only available every 60 seconds, being much less frequent than the default setup (1 sample per second). During the time where there is no GPS signal available, the trained global learning models are used to calibrate the observed sensor data. For the  i.i.d. and balanced data setup mentioned in Table~\ref{tab:setups}, we show the test performance in Fig.~\ref{fig:test-performance-fedavg-fedprox-iid}. For this case, the FedAvg algorithm is modestly superior to the FedProx algorithms in the test phase. The navigation RMSE of the FedAvg is around 9 meters, while around 12 meters for the FedProx algorithm on average. Fine-tuning the learning rate of the FedProx algorithm may further improve its generalization performance. For the non-i.i.d and imbalanced data setup shown in Table~\ref{tab:setups}, it is obvious that the FedAvg algorithm failed with a significantly degraded navigation RMSE equal to 34 meters, while the FedProx algorithm worked well with a navigation RMSE around 17 meters. We show the test performance in Fig.~\ref{fig:test-performance-fedavg-fedprox-non-iid}. For both cases, using either the FedAvg algorithm or the FedProx algorithm leads to largely improved navigation RMSE compared with 90 meters when solely using the IMU for navigation.

\section{Future Directions and Challenges}
\label{sec:challenges}
Potential challenges to the federated localization are the following:
\begin{itemize}
\item An essential ingredient of the federated wireless localization framework is the mobile terminals. To ensure that the whole framework works smoothly, the mobile terminals should be able to process a modest amount of data and perform analysis with TensorFlow, PyTorch, etc. This requires further development of powerful but compressed deep learning models, mobile AI chips, etc. Advanced WiFi and 5G technologies can fulfill the communication requirements between the mobile terminals and the central node. However, communication efficiency is a critical issue that requires more attention. In addition, an agreement on the standard protocol for synchronizing the mobile terminals is to be made. Interested readers may refer to a recent work \cite{Bonawitz2019} on how to design a scalable production system for federated learning.
\item In Section~\ref{sec:lmodels}, we mentioned that using DNN as the learning model will cause a lot of model parameters or gradients to be communicated over the air. A more straightforward and practical way to reduce the communication burden is to quantize the DNN weights from 64 bits precision to 8 bits precision or even lower. In the context of distributed optimization, a signSGD method was proposed in \cite{Bernstein2018} that quantizes every gradient update to it’s binary sign thus reducing the communication load by a factor of 32. However, better understandings on the converge properties of such methods under practical setup, such as non-i.i.d. data distribution and imbalanced data size across mobile users, need to be built.
\item The federated learning framework requires mobile users to cooperate. However, there might be the case that some voluntary mobile users are malicious or careless with their shared messages. A promising way to solve such issues from the algorithmic perspective is to use robust distributed optimization \cite{Bental09,Gorissen15}, robust estimation \cite{Zoubir2018}, and robust fusion \cite{Gustafsson2010} techniques for remedy.
%
%
\item We have so far implicitly assumed that all the mobile users have sufficient number of local data for updating the global model hyper-parameters. This may not be true for voluntary users with very limited amount of local data. One effective way to alleviate this ``small data'' difficulty from algorithmic perspective is to harness the full basket of known canonical parametric models to generate some virtual data and mix them with the small batch of real data before training the model. In this way, we are able to transfer the prior knowledge of the canonical models to our desired data-driven, learning-based model \cite{Zappone19}. 
\item We have talked exclusively about wireless localization. Actually, visual-based localization and target tracking have also attracted a lot of attention these days. The combination of wireless measurements and visual measurements can effectively improve both the localization accuracy and the robustness. For instance, in \cite{Hu19} wireless positioning was adopted in visual trackers to alleviate visual tracking pains, such as long-term tracking, feature model drifting, and recovery. Their combination is a key enabler for autonomous driving and other robotic applications. However, the inhomogeneous data structure is a big challenge to federated learning. 
\item One could utilize the social relationship of mobile users to invite more participants to join the learning process and stimulate the activeness of current participants. To this end, graph learning models, for instance graph neural network \cite{kipf2016semisupervised} and graph GP \cite{Ng18}, can be adopted for efficient learning from graph-like structured data sets.
\end{itemize}

\section{Conclusion}
\label{sec:conclusion}
In this overview paper, we reviewed all required building blocks of a fundamentally new cooperative localization and location data processing framework, called FedLoc. Being different from most of the overview papers, we put more effort on real use cases of the FedLoc framework as well as their practical implementations. We strongly believe that the FedLoc framework is promising for the following good reasons. First, high-precision wireless localization is desperately demanded, which can be achieved by combining empirical models with data driven models. Second, calibrating a localization algorithm often consumes a lot of time and workforce, and collaboration among mobile users can largely facilitate the calibration effort. Third, smartphones are becoming a powerful platform for heavy computations. Fourth, we have seen rapid development in large-scale non-convex optimization techniques, 5G communication networks, data encryption, among other emerging techniques. Lastly and most importantly, data privacy issue can be well addressed by the federated learning framework so that mobile users dare to share their location related information with safeguard.  

\section*{Acknowledgement}
We would like to thank Wenbiao Guo and Ang Xie from Beijing Jiaotong University and Haole Chen from Wuhan University for their kind help on the manuscript. This work was mainly supported by the Natural Science Foundation of China with grant No. 61701426 and partly supported by the National Key R\&D Program of China with grant No. 2018YFB1800800 and Guangdong Zhujiang Project with grant No. 2017ZT07X152.

\bibliographystyle{IEEEbib}
\bibliography{ref}

\end{document}